\documentclass[twocolumn,reprint, amssymb, amsmath,nobibnotes, aps, prb, floatfix,superscriptaddress,10pt]{revtex4-1}
\usepackage{bm}%
\usepackage{graphicx}
\usepackage[toc,page]{appendix}
\usepackage{perpage}
\usepackage{dsfont}
\usepackage{amsmath,amssymb,mathrsfs}
\usepackage{latexsym}
\usepackage{graphicx} 
\usepackage{braket}
\usepackage{epstopdf}
\usepackage{tikz}
\usetikzlibrary{matrix}
\usepackage{graphicx,epstopdf,color}
\usepackage{amsfonts}
\usepackage{amsmath,amssymb,mathrsfs}
\usepackage{hyperref}
\usepackage{amsmath} 
\usepackage{bm}
\usepackage{LatexCommands}

\allowdisplaybreaks[1]

\begin{document}

\author{Lo\"{i}c Henriet}
\affiliation{Centre de Physique Th\'{e}orique, \'{E}cole Polytechnique, CNRS, Universit\'{e} Paris-Saclay, F-91128 Palaiseau, France}
\author{Antonio Sclocchi}
\affiliation{Centre de Physique Th\'{e}orique, \'{E}cole Polytechnique, CNRS, Universit\'{e} Paris-Saclay, F-91128 Palaiseau, France}
\affiliation{Politecnico di Torino, Torino, Italy}
\author{Peter P. Orth}
\affiliation{Department of Physics and Astronomy, Iowa State University, Ames, Iowa 50011, USA}
\author{Karyn Le Hur}
\affiliation{Centre de Physique Th\'{e}orique, \'{E}cole Polytechnique, CNRS, Universit\'{e} Paris-Saclay, F-91128 Palaiseau, France}

\title{Topology of a dissipative spin: dynamical Chern number, bath induced non-adiabaticity and a quantum dynamo effect} 
\date\today
\begin{abstract}
We analyze the topological deformations of a spin-1/2 in an effective magnetic field induced by an ohmic quantum dissipative environment at zero temperature. From Bethe Ansatz results and a variational approach, we confirm that the Chern number is preserved in the delocalized phase for $\alpha<1$. We report a divergence of the Berry curvature at the equator when $\alpha_c=1$ that appears at the localization Kosterlitz-Thouless quantum phase transition in this model. Recent experiments in quantum circuits have engineered non-equilibrium protocols in time to access topological properties at equilibrium from the measure of the (quasi-)adiabatic out-of-equilibrium spin expectation values. Applying a numerically exact stochastic Schr\"{o}dinger equation we find that, for a fixed sweep velocity, the bath induces a crossover from (quasi-)adiabatic to non-adiabatic dynamical behavior when the spin bath coupling increases. We also investigate the particular regime $H/\omega_c \ll v/H \ll 1$, where the dynamical Chern number observable built from out-of-equilibrium spin expectation values vanishes at $\alpha=1/2$. In this regime, the mapping to an interacting resonance level model enables us to characterize the evolution of the dynamical Chern number in the vicinity of $\alpha=1/2$. Then, we provide an intuitive physical explanation of the breakdown of adiabaticity in analogy to the Faraday effect in electromagnetism. We demonstrate that the driving of the spin leads to the production of a large number of bosonic excitations in the bath, which in return strongly affect the spin dynamics. Finally, we quantify the spin-bath entanglement and build an analogy with an effective model at thermal equilibrium. 
\end{abstract}
\maketitle

\section{Introduction}
The notion of topology plays a key role in condensed matter systems. Applications of homotopy techniques lead, for example, to the discovery of striking hydrodynamic behaviour in superfluid helium-3~\cite{Volovik_mineyev_1,Poenaru_Toulouse, Woelfle_book} and greatly simplified the description of defects in this~\cite{Volovik_mineyev_2,Anderson_Toulouse} and other systems such as liquid crystals~\cite{ChaikinLubensky-Book, deGennes_book}. The topology of (Bloch) wavefunctions underlies the quantization of transport in quantum (spin) Hall systems and other (symmetry-protected) topological matter~\cite{TKNN,kane2005quantum}. 

\begin{figure*}[t]
\center
\includegraphics[width=.16\textwidth]{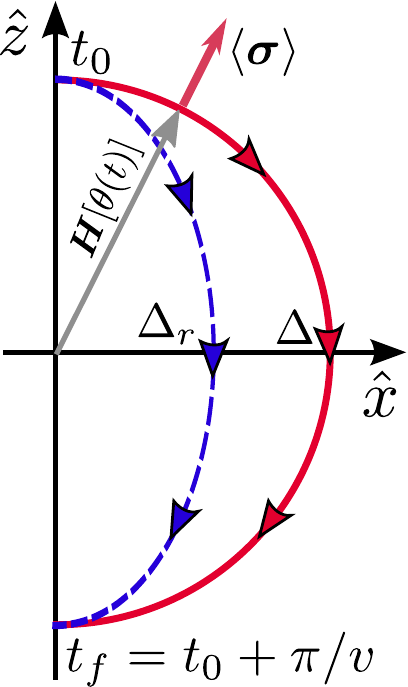}\qquad 
\includegraphics[width=.75\textwidth]{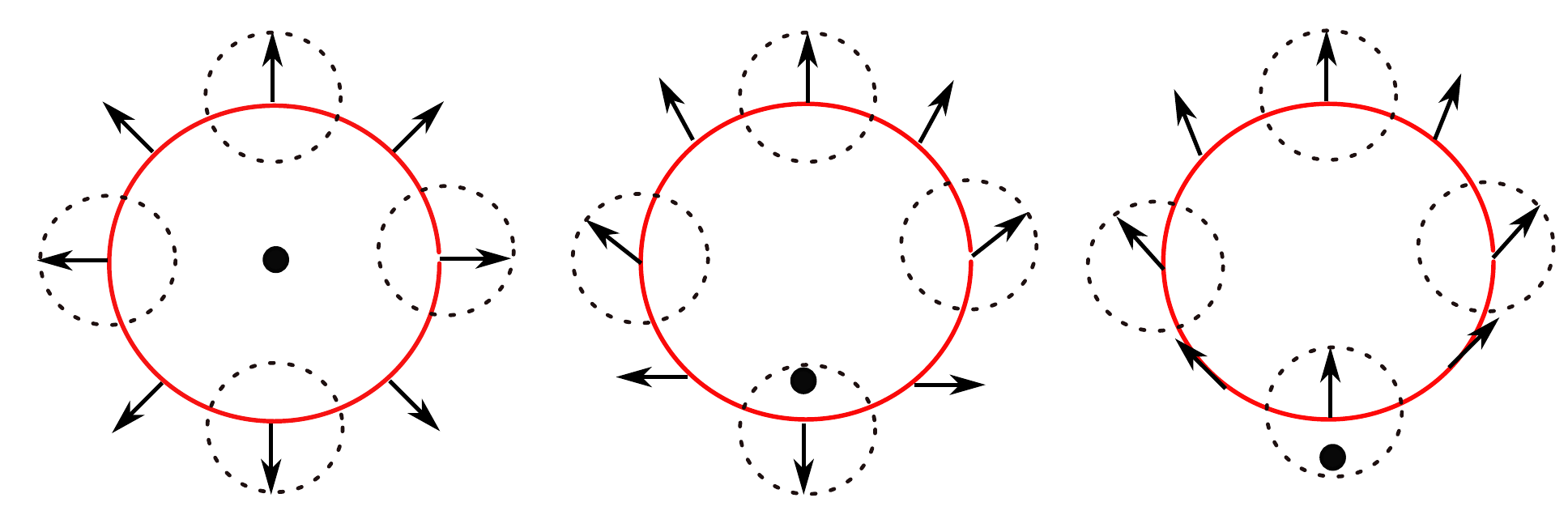}
\caption{(Left panel): Adiabatic sweep protocol of the magnetic field $\bfhh[\theta(t)]$ with linearly varying polar angle $\theta = v (t - t_0)$. Grey arrow denotes external magnetic field, magenta arrow the spin direction in the ground state. Red path is for a free spin (or weak spin-bath coupling), blue bath is for stronger spin-bath coupling which leads to renormalization of transvere field $\Delta = H \sin \theta \rightarrow \Delta_r$. (Three right panels): The red circles are parametrized by $(H \sin \theta,H_0+H \cos \theta)$ with $H\neq 0$. We have $H_0/H=0$ (left), $0<H_0/H<1$ (middle) and $H_0/H>1$ (right) and the black dot shows the position of the origin in each case. The arrows show the orientation of the Bloch vector for each value of $\theta$. The vertical component is given by $\langle \sigma^z \rangle_{eq}$ while the horizontal component is given by $\langle \sigma^x \rangle_{eq}$.}
\label{topo_TLS_isole}
\end{figure*}

To illustrate the notion of topology in a simple example, let us consider a quantum spin-$1/2$ particle in a magnetic field, as described by the Hamiltonian
\begin{align}
\mathcal{H}_{TLS}= - \frac{1}{2} \boldsymbol{h} \cdot \boldsymbol{\sigma},
\label{Htls_topo}
\end{align}
where $\boldsymbol{h} = \bfhh + \bfhh_0$ describes a magnetic field with is a superposition of a field $\bfhh$ in the radial direction $\bfhh = (H \sin \theta \cos \phi , H \sin \theta \sin \phi, H \cos \theta )$ and a constant field along $\hat{z}$: $\bfhh_0 = (0,0,H_0)$. Here, $\theta$ and $\phi$ represent the polar and azimuthal angles on the sphere, and $\boldsymbol{\sigma} = (\sigma^x, \sigma^y, \sigma^z)$ is a vector of Pauli matrices. For $\phi=0$, the ground state Bloch vector $\av{\boldsymbol{\sigma} }= \braopket{g}{\boldsymbol{\sigma} }{g}$ aligns with the direction of $\bfh(\theta, \phi)$ and lies on a unit circle in the $(xOz)$ plane, where $\ket{g}$ denotes the ground state. In Fig. \ref{topo_TLS_isole}, we show the orientation of $\av{\boldsymbol{\sigma}}$ for $\theta \in [0,2\pi)$ and $\phi =0 $. The winding behaviour of the spin around the circle depends on the ratio $H_0/H$. For $H_0/H < 1$, the angle between the Bloch vector and the vertical $\hat{z}$ axis runs from $0$ to $2 \pi$ when $\theta$ is changed in the same range. In contrast, this angle comes back to zero for $H_0/H > 1$. More generally, since the Bloch vector $\av{\boldsymbol{\sigma}}$ must be identical at $\theta=0$ and $\theta=2 \pi$ for any $2\pi$-periodic Hamiltonian, it winds an integer number of times around $2 \pi$ as $\theta$ changes from zero to $2 \pi$. This topological (Chern) winding number $C \in \mathbb{Z}$ is a characteric of the ground state $\ket{g}$ of Hamiltonian~(\ref{Htls_topo}), and we find $C = 1$ for $H_0 < H$, while $C = 0$ for $H_0>H$ (see Appendix~\ref{appendix_index} for more details). 

While the Chern number $C$ is a global topological property of the state $\ket{g}$, one can also define quantities describing the local topology (or geometry). This information is contained in the Berry curvature $\mathcal{F}_{\phi\theta }$~\cite{Berry}. It characterizes the local geometry of a state $\ket{g}$ upon infinitesimal variation of the angles $\theta$ and $\phi$, and is defined by 
\begin{align}
\mathcal{F}_{\phi\theta}=\partial_{\phi} \mathcal{A}_{\theta}-\partial_{\theta} \mathcal{A}_{\phi},
\label{Berry_curvature}
\end{align}
where $\mathcal{A}_{\phi}$ and $\mathcal{A}_{\theta}$ are the Berry connections defined by ($\alpha = \theta, \phi$)
\begin{align}
\mathcal{A}_{\alpha}=\langle g|i\partial_{\alpha}|g \rangle \,.
\label{Berry_connection}
\end{align}
The Chern number $C$ and the Berry curvature $\mathcal{F}_{\theta \phi}$ are gauge-independent quantities, whereas $\mathcal{A}_\alpha$ depends on the gauge phase of state $\ket{g}$. For the Hamiltonian in Eq.~\eqref{Htls_topo} with $H_0 = 0$, we find $\mathcal{F}_{\phi \theta}= \frac12  \sin \theta$ for the ground state $|g\rangle$. The Chern number $C$ is obtained from integrating the curvature over the full Bloch sphere
\begin{align}
\label{eq:3}
C=\frac{1}{2\pi}\int_{0}^{2\pi} d \phi \int_{0}^{\pi} d \theta \mathcal{F}_{\phi \theta}.
\end{align}
For $H_0 = 0$, we thus find $C=1$. Calculating $\mathcal{F}_{\theta \phi}$ using the ground state wavefunction for non-zero $H_0$, one can easily show that the Chern number remains $C=1$ as long as $H_0/H < 1$ and jumps to $C=0$ for $H_0/H>1$. The winding properties of the spin can also be accessed using an adiabatic sweep protocol of slowly varying the polar angle ($v/H \ll 1$)
\begin{equation}
  \label{eq:1}
  \theta(t)=v(t-t_0)
\end{equation}
with $t \in [t_0, t_0 + \pi/v]$. This sweep protocol is illustrated in Fig.~\ref{topo_TLS_isole}. Solving the spin dynamics using Heisenberg equations of motion for this sweep in the precise case of Hamiltonian (\ref{Htls_topo}) where $C\in\{0,1\}$, one can explicitly derive that the Chern number is determined by the direction of the Bloch vector at the north and south pole along the path and given by 
\begin{align}
C=\frac{\langle \sigma^z(\theta=0) \rangle-\langle \sigma^z(\theta=\pi) \rangle}{2} \,.
\label{C_sigma_z}
\end{align}

While we illustrated the notions of Berry curvature, Berry connection and Chern numbers using this simple example of a single isolated spin in a magnetic field, these definitions are very general and non-trivial topological effects can arise for any Hamiltonian that is periodic in a certain variable. Most importantly, electrons in a crystal can be described by a Bloch Hamiltonian that is periodic in electronic momentum $\bfk$, \emph{i.e}, it is invariant under adding a reciprocal lattice vector to $\bfk$. In this context, one may see the Hamiltonian in Eq.~(\ref{Htls_topo}) as an electronic single-particle Hamiltonian of the simplest class of Chern insulators with only two bands, which notably includes the well-known Haldane model\cite{haldane1988model}. The physical significance of a non-zero Chern number $C_\alpha$ of Bloch band $\alpha$ is that it may lead to a quantized Hall conductivity $\sigma_{xy}=\frac{e^2}{2\pi \hbar}\sum_{\alpha} C_{\alpha}$ where the sum runs over all the filled bands below the Fermi level~\cite{TKNN}. 

Returning to the topology of a Bloch spin, it is important to realize that most physical systems are in practice not completely separated from their environment. This leads to the phenomenon of dissipation, which can classically and phenomenologically be described by a smooth frictional force that accounts for the loss of energy and a fluctuating force (of zero average) that accounts for the randomness of the energy exchange~\cite{weiss}. These two forces are closely related to each other by the fluctuation-dissipation theorem. Classically, the environment is a heat reservoir at a certain temperature $T$ which constitutes a noise source. If the reservoir exhibits time retardation effects, its noise is described as colored, and a friction force that does not depend on the state of the system can then be captured by a linear functional that takes into account the motion of the system at previous times through a non-Markovian memory friction kernel.

With the aim of recovering this description in the classical limit, various approaches to open quantum systems were introduced~\cite{weiss}. The most successful one has been to consider coupling the system of interest to an environment with infinitely many degrees of freedom. If the state of the environment is only weakly perturbed by the coupling to the system, the environmental degrees of freedom can be described by an infinite set of harmonic oscillators. This scheme defines the class of Caldeira-Leggett models~\cite{leggett1981,leggett:RMP}, leading to a microscopic Hamiltonian $\mathcal{H}=\mathcal{H}_{TLS}+\mathcal{H}_{diss}$ with 
\begin{align}
\mathcal{H}_{diss}&= \sigma^z \sum_k \frac{\lambda_k}{2} (b_k +b_k^{\dagger})+ \sum_k \omega_k \left(b_k^{\dagger}b_k+\frac{1}{2}\right).
\label{Hamiltonian_spin_boson} 
\end{align}
Here, $b_k^{\dagger}$ is the creation operator of a boson in mode $k$ with frequency $\omega_k$ (we set the Planck constant $\hbar=1$). The spin-bath interaction is fully characterized by the spectral function $J(\omega)=\pi \sum_k \lambda_k^2 \delta(\omega-\omega_k)$, which we assume to be of Ohmic form 
\begin{equation}
  \label{eq:2}
  J(\omega) = 2 \pi \alpha \omega \exp \left(-\frac{\omega}{\omega_c} \right) \,.
\end{equation}
Here, $\alpha$ describes a dimensionless dissipation strength and $\omega_c \gg H$ denotes a high energy bath cutoff energy, which is the largest energy scale in the problem. An environment with Ohmic spectral density is a valid description for a number of different systems, for example, for in circuit quantum electrodynamics (cQED) where the environment is embodied by long transmission lines~\cite{Cedraschi:PRL,Cedraschi:Annals_of_physics}, in cold atomic setups with one-dimensional Bose-Einstein condensates\cite{recati_fedichev}, or in Luttinger liquids~\cite{InesSaleur,KLH2}. 

The Ohmic spin-boson model is known to exhibit a dissipative quantum phase transition at $\alpha_c = 1$ that separates a delocalized phase, where the spin expectation value $\av{\sigma^z}$ vanishes in the absence of a bias field along the $\sigma^z$, from a (symmetry-broken) localized phase where $\langle\sigma^z\rangle \neq 0$ already for infinitesimal bias field along $\sigma^z$. In addition, this model shows a coherent-to-incoherent crossover at $\alpha = 1/2$ in the dynamical Rabi-type properties of the spin dynamics~\cite{leggett:RMP, weiss,PhysRevB.88.165133}. We note that the spin-boson model is intimately related to a one-dimensional Ising model with long-range interactions and to the Kondo model~\cite{Anderson_Yuval_Hamann,Blume_Emery_Luther,Karyn}.

The presence of a bath naturally leads to the interesting question whether a coupling to the dissipative environment affects the topology of the spin. For strong spin-bath coupling above the critical coupling strength, \emph{i.e.}, $\alpha > \alpha_c = 1$, the spin is localized~\cite{leggett:RMP,weiss} and tunneling between $\ket{\uparrow_z}$ and $\ket{\downarrow_z}$ eigenstates of $\sigma^z$ do not occur even in the presence of a transverse field term $H \sin (\theta) \sigma^x = \Delta \sigma^x$. The spin is trapped in a polarized state along the $\hat{z}$-axis (even for small fields along $\sigma^z$). The possible equilibrium Bloch vectors in the localized phase are not connected by a continuous path on the Bloch sphere. Instead, the expectation value of the spin discontinuously jumps from being $\av{\sigma^z} = 1$ (for $H_z = H \cos \theta > 0$) to being $\av{\sigma^z} = -1$ for $H_z < 0$, and the Chern number is no longer well-defined. At sufficiently weak coupling $\alpha \ll 1$, on the other hand, one expects that the bath cannot change the global topology of the spin, expressed by the Chern number $C$. The coupling to the bath, however, may affect the local geometry of the spin, which is described by the Berry curvature $\mathcal{F}_{\phi \theta}$. Recent experiments in circuit QED~\cite{Haeberlein_arxiv,Forn_diaz} have realized the large coupling limit for an Ohmic bath. It is thus interesting to investigate whether a possible bath-induced geometrical deformation of the Berry curvature can be accessed experimentally. 

In the remainder of this article, we investigate the bath induced changes of the spin topology and how it may be accessed experimentally. We study the topology both in the ground state and within a recently proposed (almost) adiabatic dynamical sweep protocol that is relevant experimentally~\cite{polkovnikov:PNAS,Roushan:Nature,schroer2014measuring,Schroer:PRL}. 
In Sec.~\ref{sec:robustn-equil-chern}, we access the topological properties of the ground state both using results from the exact Bethe ansatz and an approximate variational approach. 
In Sec.~\ref{results_spin_dynamics}, we address the question how to measure the Berry curvature and Chern number using a dynamic sweep protocol characterized by a small frequency $v/H \ll 1$. We compute the spin dynamics and the dynamical Chern number $C_{\text{dyn}}$ using the numerically exact Stochastic Schr\"{o}dinger Equation (SSE) technique, which fully accounts for the non-Markovian effects of an Ohmic bath at low temperature. We demonstrate that as the spin-bath coupling $\alpha$ is increased, the bath inevitably induces a crossover from (quasi-)adiabatic to non-adiabatic dynamical behavior during the sweep. As a result, the dynamically measured Chern number $C_{\text{dyn}}$ deviates from the ground state result $C$ already for $\alpha < \alpha_c = 1$. In a sense, the equilibrium topological properties are screened by the environment. For fixed velocity $v/H \ll 1$ we find that the crossover to non-adiabatic behavior occurs when $v/\Delta_r \approx 1$, where $\Delta_r = \Delta (\Delta/\omega_c)^{\alpha/(1- \alpha)} < \Delta$ is the bath renormalized transverse field. We the focus on the experimentally relevant regime of $H > v >  \Delta_r$ relevant to baths with a large bandwidths $\omega_c \gg H$. We observe that for fixed velocities in this range, the dynamic Chern number $C_{\text{dyn}} \rightarrow 0$ as $\alpha \rightarrow 1/2$ in the universal scaling regime $\omega_c \rightarrow \infty$ with fixed $\Delta_r < v$. We analytically determine how $C_{\text{dyn}}$ approaches zero using an exact mapping to the non-interacting resonance level model (which is exactly soluble at the Toulouse point $\alpha = 1/2$). 

In Sec.~\ref{sec:radi-casc-phot} we use a toy model to provide an intuitive physical interpretation of this breakdown of $C_{\text{dyn}}$ in terms of a resonant excitation of bath modes, which we term the ``quantum dynamo effect''. 
Finally, in Sec.~\ref{sec:entangl-entr-effect} we study the evolution of the entanglement entropy due to spin-bath coupling and introduce an effective thermodynamical description of the quantum dynamo effect. 
We conclude in Sec.~\ref{experimental_realizations} and present an outlook on various experimental setups that may be able to access the spin topology that we describe. We provide a number of calculational details in the Appendices. 

\section{Equilibrium Chern number of a dissipative spin}
\label{sec:robustn-equil-chern}
To explore the effect of the bath on the geometrical properties of the spin for general coupling strength $\alpha$, let us express the ground state wavefunction in the general form 
\begin{align}
\ket{g} =\frac{1}{\sqrt{p^2+q^2}}\left[p e^{-i \phi} \ket{\uparrow_z} \otimes \ket{\chi_{\uparrow}} + q \ket{\downarrow_z} \otimes \ket{\chi_{\downarrow}} \right]\,.
\label{ansatz}
\end{align}
Here, $p$ and $q$ are two real numbers and $\ket{\chi_{\sigma}}$ is a bath state associated with the spin polarization $\sigma = \uparrow, \downarrow$. Due to the symmetry in the Hamiltonian, which does not contain $\sigma^y$, these quantities only depend on the polar angle $\theta$, but are independent of the azimuth $\phi$, \emph{i.e.}, $p=p(\theta), q=q(\theta), \ket{\chi_\sigma} = \ket{\chi_\sigma(\theta)}$. The only dependence of $\ket{g}$ on $\phi$ is via the phase factor in Eq.~(\ref{ansatz}). Using Eq.~\eqref{ansatz} we find a general expression of the Berry connections $\mathcal{A}_\alpha$ and the curvature $\mathcal{F}_{\theta \phi}$ in terms of $p$ and $q$. We find that $\mathcal{A}_{\theta}$ does not depend on $\phi$ and thus $\partial_\phi \mathcal{A}_\theta = 0$. The connection associated with $\phi$ is given by
\begin{align}
&\mathcal{A}_{\phi}=\langle g|i\partial_{\phi}|g \rangle=\frac{p^2}{p^2+q^2}\,.
\label{Berry_connection_ansatz}
\end{align}
The Berry curvature follows from Eq.~(\ref{Berry_curvature}) as $\mathcal{F}_{\phi \theta} = - \partial_\theta \mathcal{A}_\phi = - \partial_\theta [p^2/(p^2 + q^2)]$. Interestingly, this allows to derive a useful relation between the Berry curvature and the spin susceptibility with respect to $\theta$. Using that $\av{\sigma^z} = (p^2-q^2)/(p^2+q^2)$, one finds
\begin{align}
&\mathcal{F}_{\phi \theta}=-\partial_{\theta} \langle \sigma^z \rangle/2.
\label{Berry_connection_sigma_z}
\end{align}
While the coupled spin-bath ground state can generally be written in the form of Eq.~(\ref{ansatz}), it is far from trivial to compute the coefficients $p$ and $q$ as well as bath states $\ket{\chi_{\sigma}}$ for a given value of $\theta$. 

Physically, the presence of the bath tends to increase the polarization of the spin along the axis parallel to the spin compared to the isolated case. This can be easily understood for the case of a completely polarized spin (\emph{i.e.} for large external bias fields $h_z$). A bath that has equilibrated with a completely polarized spin along direction $\sigma = \pm 1$ is in a shifted bath oscillator state as described by the bath reduced density matrix $\rho_B(\sigma) = \frac{1}{Z} \exp[ - \beta \sum_k (\omega_k b^\dag_k b_k + \frac{\sigma}{2} \lambda_k (b^\dag_k + b_k))]$ with $\beta = 1/T$ and $Z = \text{Tr} \rho_B$. In that state, the bath acts in turn on the spin as an effective magnetic field along the $\sigma^z$ direction due to the term $\sigma^z \sum_{k} \lambda_k (b^\dag_k + b_k)$ in the Hamiltonian~\cite{weiss, stochastic}. While the strength of the effective field $h_{B,z}$ is reduced for a spin that is only partially polarized, a non-zero magnetization of the spin $\av{\sigma^z} > 0$ imposed by a magnetic field with $H_z > 0$, \emph{i.e.} $\theta < \pi/2$, nevertheless results in a polarization of the bath. The resulting effective field $h_{B,z}$ adds to the external field $h_z$ resulting in an increased spin polarization compared to the case of a free spin.

\begin{figure}[ht]
\center
\includegraphics[width=\linewidth]{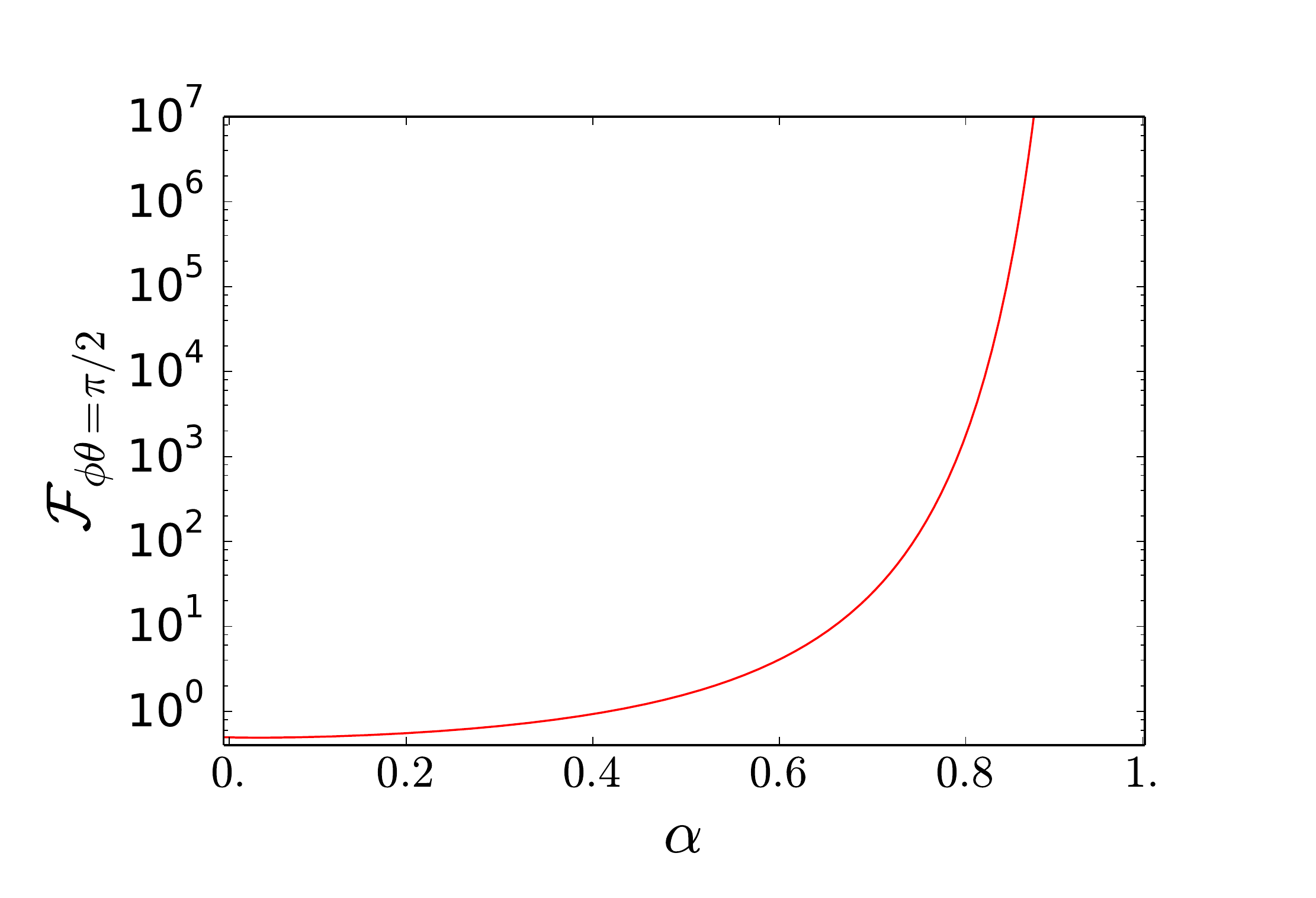}
\includegraphics[width=\linewidth]{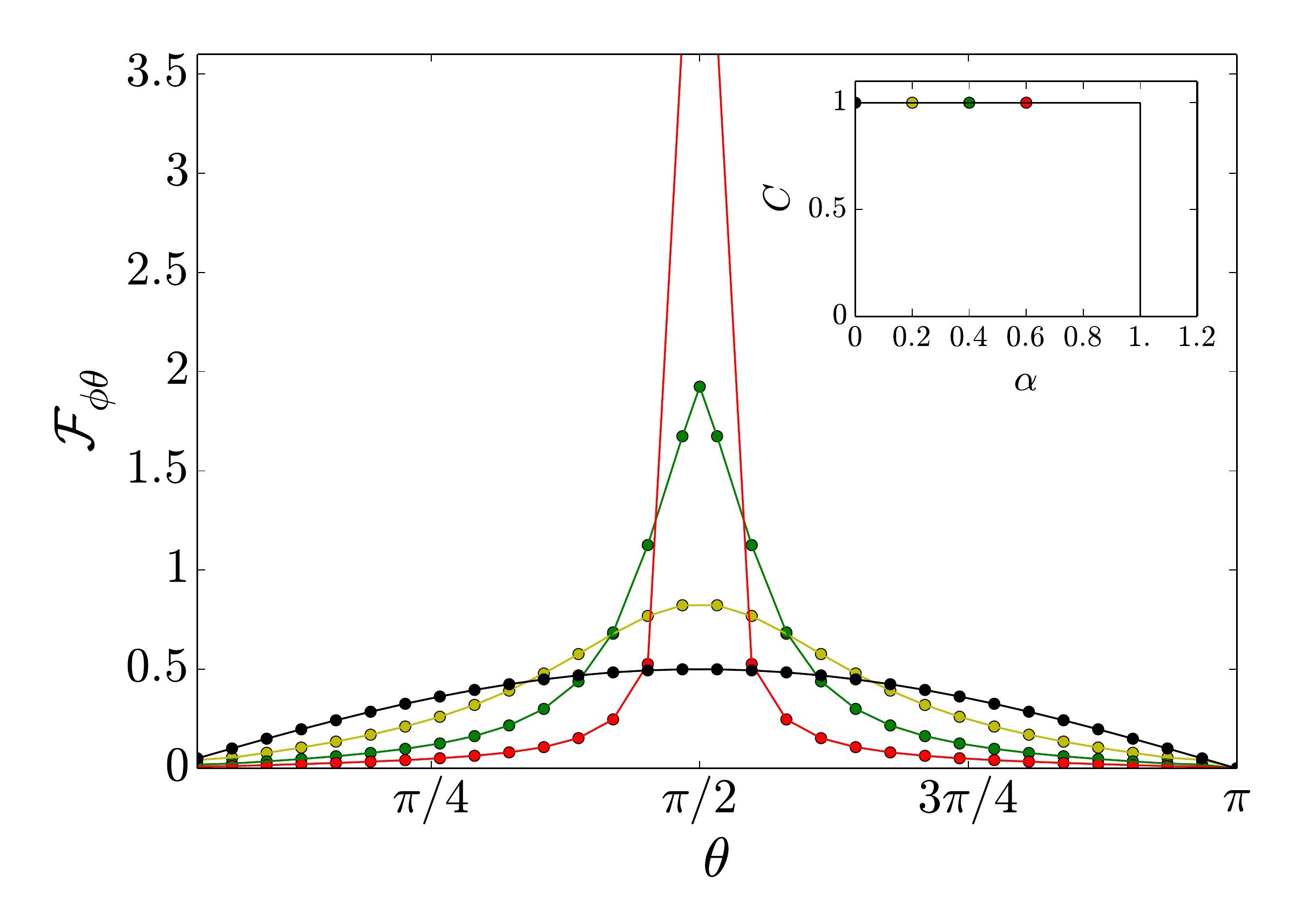}
\caption{(Color online) (Upper panel) Evolution of the Berry curvature at the equator $\mathcal{F}_{\phi \theta=\pi/2}$ from Bethe Ansatz results, Eq. (\ref{F_bethe_equator}) for $H/\omega_c=0.2$, allowing to access the vicinity of the quantum phase transition at $\alpha_c=1$ at equilibrium. (Lower panel) Evolution of the Berry curvature $\mathcal{F}_{\phi \theta}$ in the single polaron picture obtained with a variational approach valid at low coupling, with respect to $\theta$ for $\alpha=0$ (black), $\alpha=0.2$ (yellow), $\alpha=0.4$ (green), $\alpha=0.6$ (red). We have chosen the particular value $H/\omega_c=0.2$. The inset shows the evolution of the Chern number with $\alpha$. The dots are obtained from the integration of the Berry curvature obtained from the single-polaron ansatz.}
\label{Berry_curvature_shifted_oscillators}
\end{figure}

The ability of the bath to influence the spin direction in the ground state affects the topology of the spin. Using the powerful Eq.~\eqref{Berry_connection_sigma_z}, we can employ exact Bethe ansatz results for the spin expectation values \cite{Bethe_1,Bethe_2,Hur} to determine analytically the evolution of the Berry curvature $\mathcal{F}_{\theta \phi}$. Most interestingly, inserting the exact expression for $\langle \sigma^z \rangle$ given in Ref.~\onlinecite{Hur} into Eq. (\ref{Berry_connection_sigma_z}), we derive an exact expression for the Berry curvature at the equator $\theta = \pi/2$ valid in the full range $0<\alpha<1$:
\begin{align}
\mathcal{F}_{\phi \theta=\pi/2}=F(\alpha) \left(\frac{\omega_c}{H}\right)^{\frac{\alpha}{1-\alpha}} = F(\alpha) \frac{\Delta}{\Delta_r}\,,
\label{F_bethe_equator}
\end{align}
where we have defined the transverse field $\Delta = H \sin \theta$ (note that $\Delta = H$ at the equator $\theta = \pi/2$), the renormalized field $\Delta_r = \Delta (\Delta/\omega_c)^{\alpha/(1-\alpha)}$ and the function $F(\alpha)=1/\sqrt{\pi}(4/\pi)^{\alpha/(1-\alpha)}\exp\{b/[2(1-\alpha)]\} \Gamma[1+1/(2-2\alpha)]/\Gamma[1+\alpha/(2-2\alpha)] $. Here, $\Gamma$ denotes the incomplete Gamma function and $b=\alpha \ln \alpha - (1-\alpha) \ln (1-\alpha)$. 

Since $H < \omega_c$, we find that the Berry curvature $\mathcal{F}_{\phi\theta}$ diverges at the equator $\theta = \pi/2$ when $\alpha \rightarrow \alpha_c = 1$, \emph{i.e.}, at the delocalized-localized quantum phase transition. This follows from the fact that $\Delta_r$ goes to zero at the phase transition, which is in the Kosterlitz-Thouless universality class~\cite{leggett:RMP,weiss,Hur}. The divergence of $\mathcal{F}_{\phi\theta = \pi/2}$ is illustrated in Fig.~\ref{Berry_curvature_shifted_oscillators}. At this critical value, the equilibrium Bloch vector manifold splits into two separate parts which are no longer connected, reflecting the jump of the order parameter $\langle\sigma^z\rangle$ at the transition \cite{Karyn}. 

We have also computed the evolution of the Berry curvature $\mathcal{F}_{\phi \theta}$ as a function of $0 \leq \alpha < 1$ using an approximate, variational approach to determine the coefficients $p(\theta)$ and $q(\theta)$ in Eq.~\eqref{ansatz}. As shown in detail in Appendix~\ref{Appendix_polaron}, we expand the bath wavefunction $\ket{\chi_\sigma}$ in terms of classical bath polaron states~\cite{Silbey_Harris, bera:PRB} and the coefficients $p,q$ are found from minimizing the energy. For simplicity, we restrict to a single polaron expansion (so-called Silbey-Harris approach~\cite{Silbey_Harris}) and show the resulting Berry curvature $\mathcal{F}_{\phi \theta}$ as a function of $\theta$ for different values of $\alpha < 1$ in Fig.~\ref{Berry_curvature_shifted_oscillators}. We observe that the environment gradually deforms the ground state manifold upon a variation of $\theta$. As expected from the exact Bethe ansatz result in Eq.~\eqref{F_bethe_equator}, the deformation is most pronounced around the equator, where the Berry curvature becomes more and more peaked as $\alpha$ increases. Importantly, however, the value of the Chern number $C$ obtained from integrating the Berry curvature over the complete Bloch sphere (see Eq.~\eqref{eq:3}) remains unchanged and equals unity for all values of $\alpha < 1$. Dissipation does not change the global topology at not too strong couplings $\alpha < 1$. This weak-coupling behavior should be quite general and also hold for other types of environments such as non-Ohmic spectral densities $J(\omega) \propto \omega^s$ with $s \neq 1$ or fermionic baths. As long as the system-environment coupling is sufficiently weak, the global topology of the spin is protected, and the Berry curvature is only locally modified.

\section{Geometrical and topological observables in a time-dependent framework}
\label{results_spin_dynamics}

\begin{figure*}[t]
\includegraphics[width=.3\textwidth]{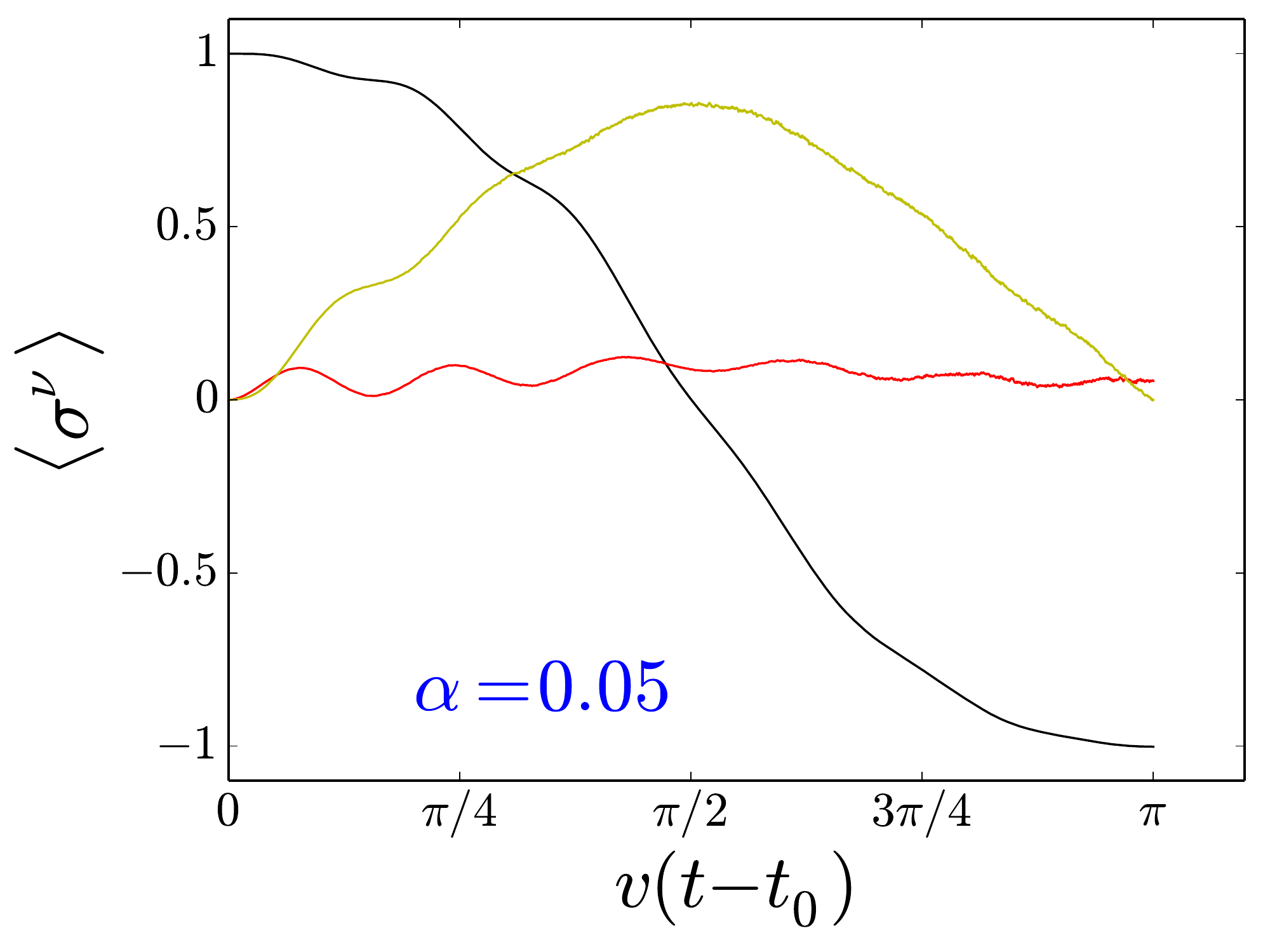}
\includegraphics[width=.3\textwidth]{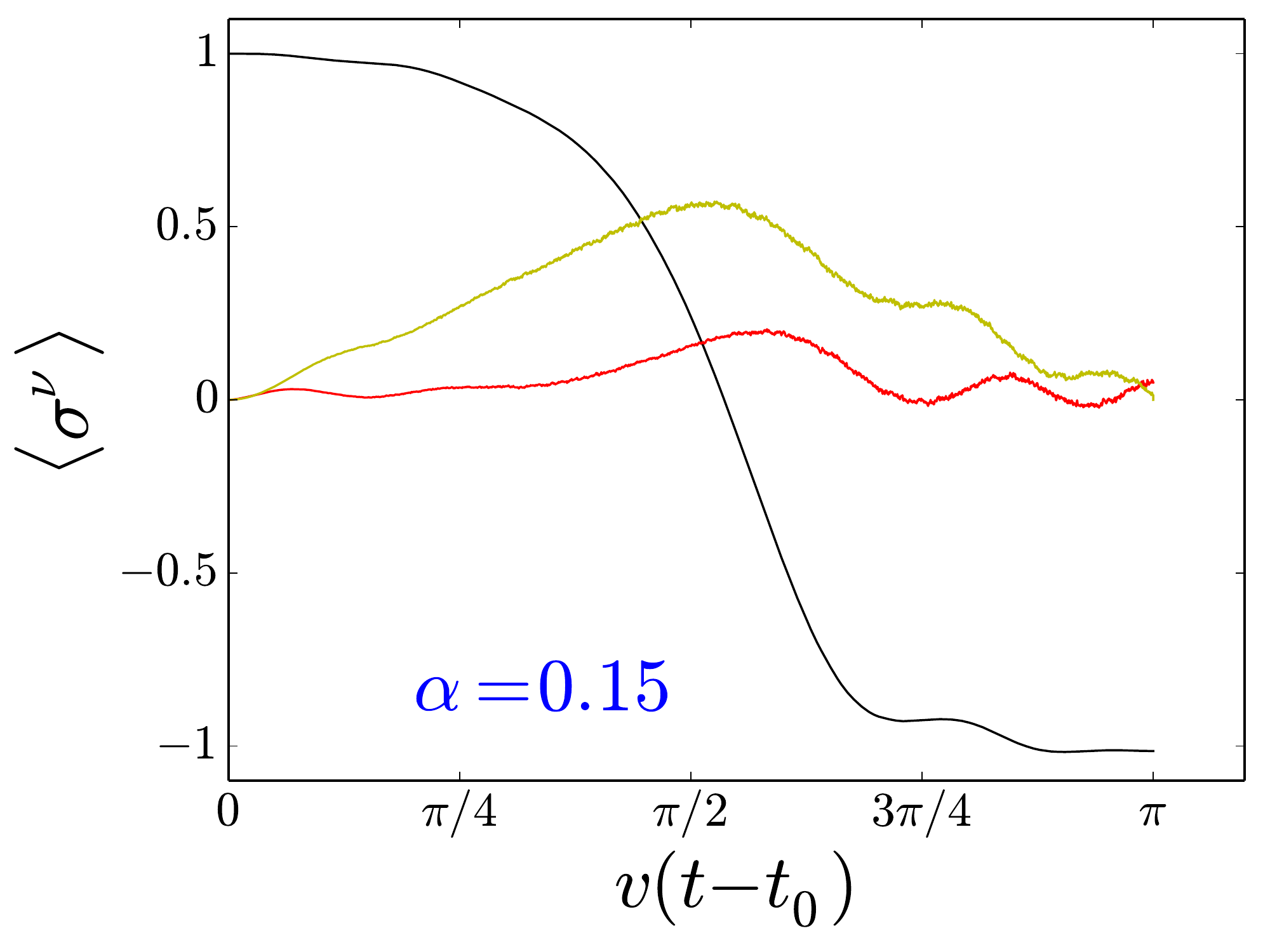}
\includegraphics[width=.3\textwidth]{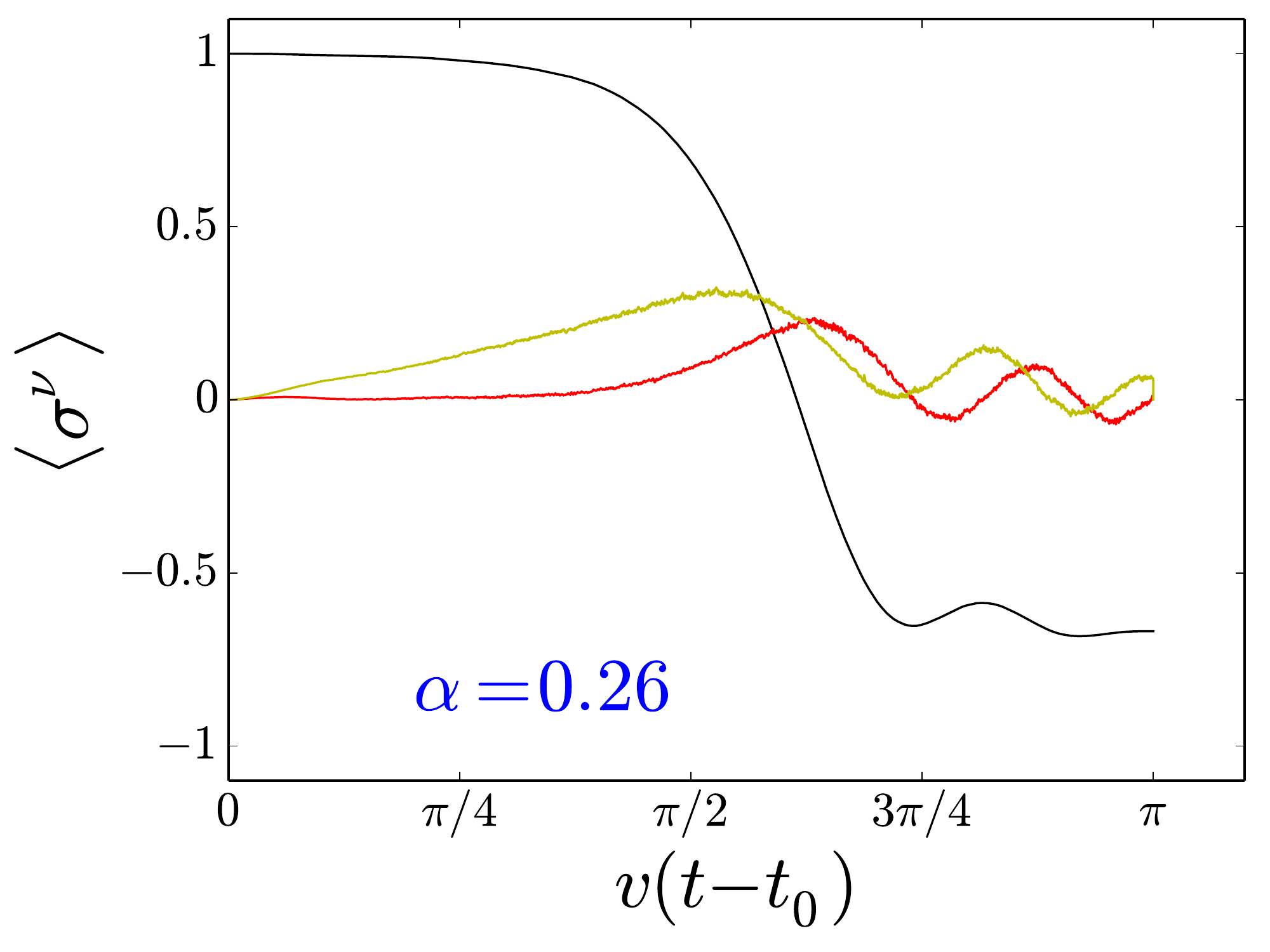}\\
\includegraphics[width=.3\textwidth]{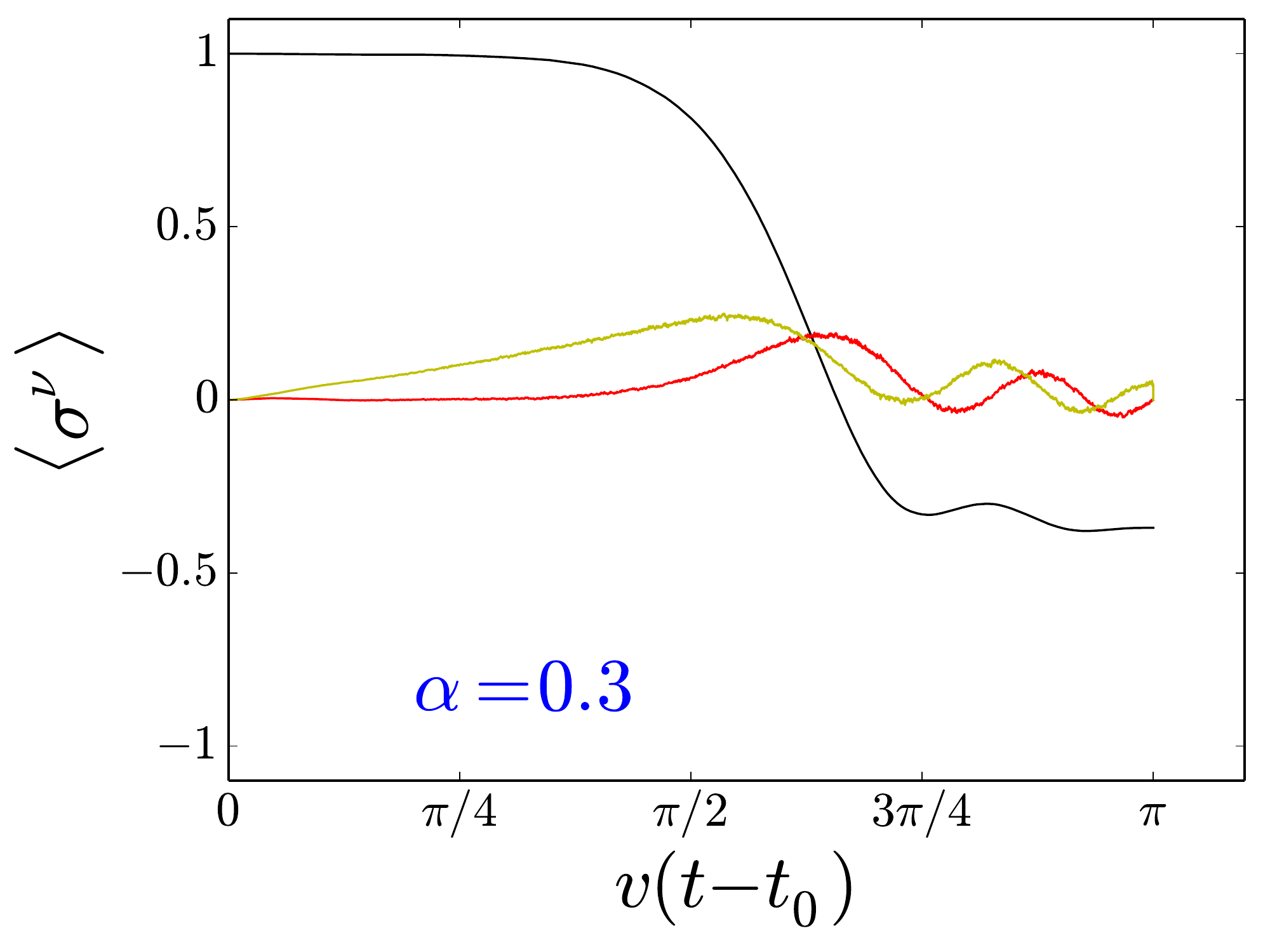}\includegraphics[width=.3\textwidth]{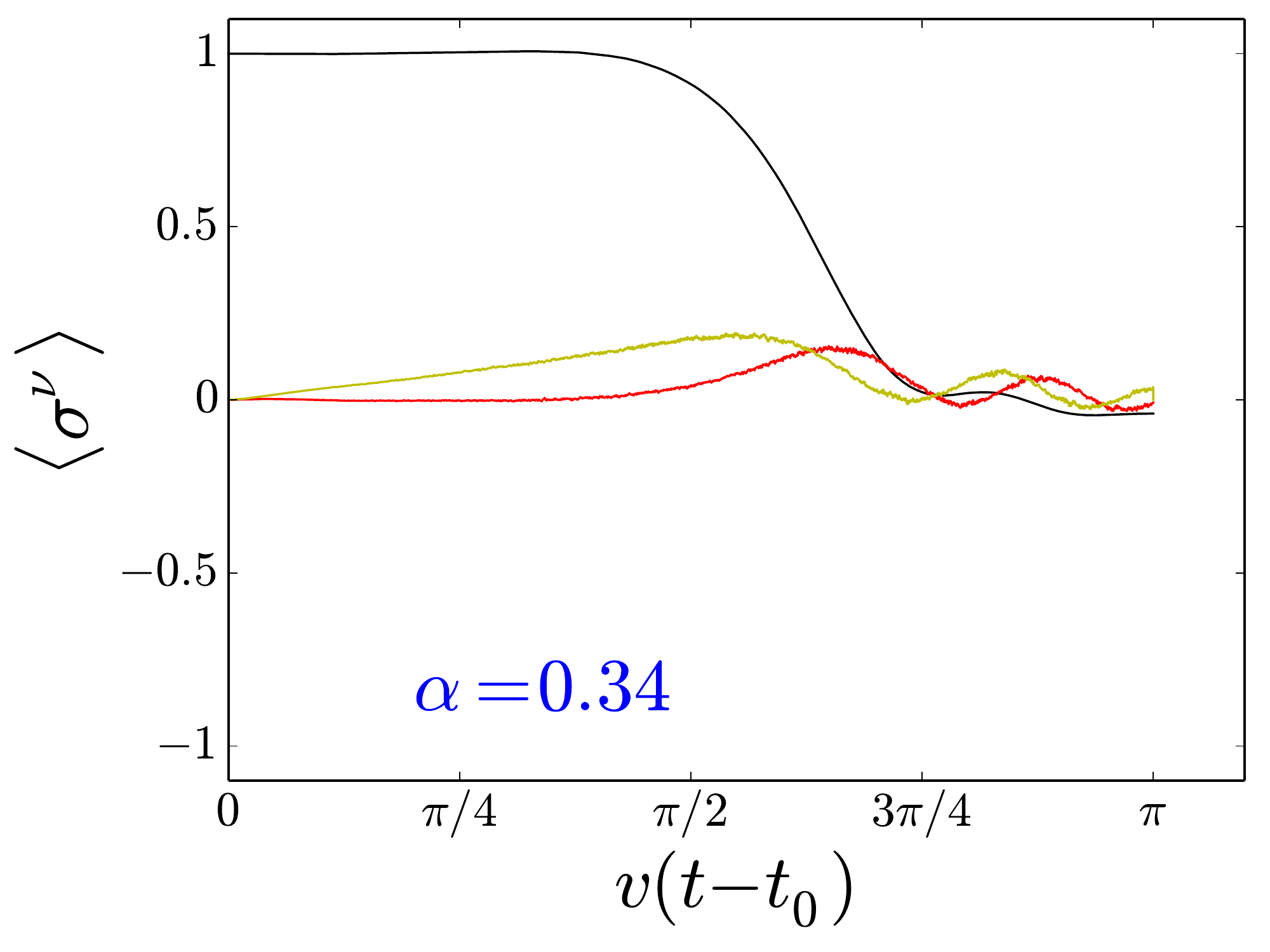}
\includegraphics[width=.3\textwidth]{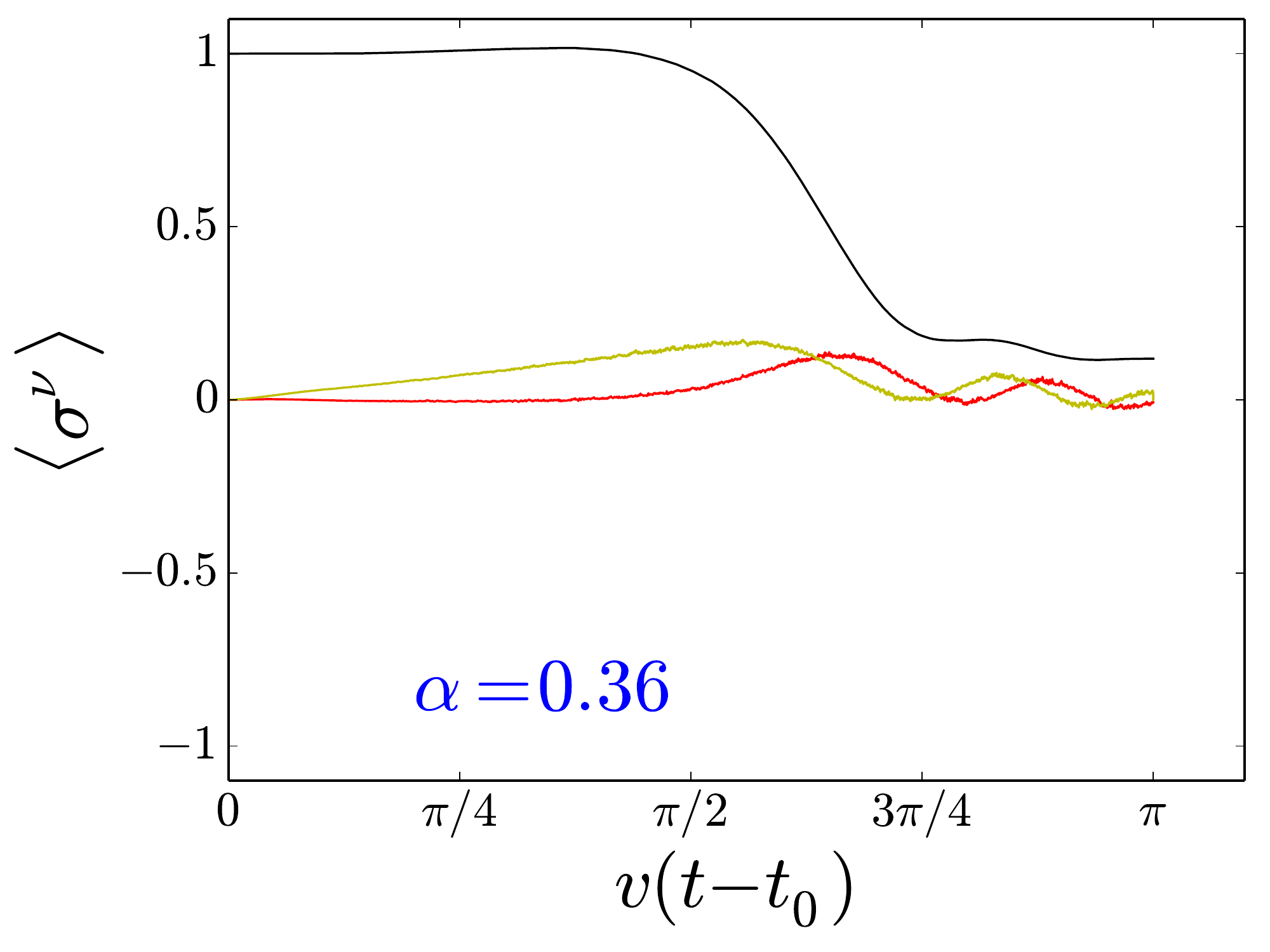}
\caption{Spin variables $\langle \sigma^z \rangle$ (black), $\langle \sigma^x \rangle$ (yellow) and $\langle \sigma^y \rangle$ (red) and their time-evolution during the sweep, with $v/H=0.08$. From left to right and top to bottom, we take $\alpha=0.05$, $\alpha=0.15$, $\alpha=0.26$, $\alpha=0.3$, $\alpha=0.34$ and $\alpha=0.36$. We have $H/\omega_c=0.01$. In the low coupling regime $\alpha \ll 1$ we recover the two different frequencies for the dynamics $v$ and $\sqrt{H^2+v^2}$, as shown by the description in the rotating frame (Eq. (\ref{rotated_basis}) and discussion below).
}
\label{dynamics_topo}
\end{figure*}

\subsection{Sweep protocol and dynamical Chern number}
\label{sec:prot-equil-dynam}  
In a recent work of Polkovnikov and Gritsev in Ref.~\onlinecite{polkovnikov:PNAS}, it was proposed that the Berry curvature $\mathcal{F}_{\phi \theta}$ of a spin can be measured via a dynamical sweep protocol. Let us first discuss this protocol for a free spin before investigating the effect of the coupling to a bath. Within the sweep protocol the spin is subject to the Hamiltonian $\mathcal{H}_{TLS}$ in Eq.~(\ref{Htls_topo}) with a time-dependent vector $\bfh(t)$ whose polar angle is changed linearly in time according to 
\begin{equation}
  \label{eq:4}
  \theta(t)=v(t-t_0) \,.
\end{equation}
Within the time interval $t \in [t_0, t_0 + \pi/v]$, the vector $\bfh(t)$ follows a half circle from the north to the south pole of the sphere. For simplicity we choose $\phi=0$ in the following. The sweep protocol is illustrated in Fig.~\ref{topo_TLS_isole}. The spin is initially prepared in the ground state $\ket{\uparrow_z}$ at time $t_0$. If the sweep velocity is small $v/H \ll 1$, the dynamics is nearly adiabatic. In the non-dissipative case $\alpha=0$, the Bloch vector spirals around the field, following a characteristic cyclo\"{i}d curve from north to south pole that is associated with oscillations of $\av{\sigma^{\alpha}}$. The dynamics can be conveniently described in a rotating frame, defined by the unitary transformation $U(t) = \exp( \frac{i}{2} \theta(t) \sigma^y)$ corresponding to a rotation around the $\hat{y}$-axis by an angle $\theta(t) = v(t-t_0)$. The state transforms according to $\ket{\psi'(t)}=U(t)\ket{\psi(t)}$ such that it obeys the Schr\"{o}dinger equation 
 \begin{align}
&i \partial_t |\psi'(t)\rangle=\left[i \dot{U}U^{\dagger}+U \mathcal{H}_{TLS}U^{\dagger}   \right]|\psi'(t)\rangle = \mathcal{H}_{\text{eff}} \ket{\psi'(t)} \,,
\label{rotated_basis}
\end{align}
with effective Hamiltonian
\begin{equation}
  \label{eq:5}
 \mathcal{H}_{\text{eff} } = i \dot{U}U^{\dagger}+U \mathcal{H}_{TLS}U^{\dagger}  = -H \tilde{\sigma}^z/2 -v \tilde{\sigma}^y/2 \,.
\end{equation}
The spin operators $\tilde{\sigma}^\alpha$ refer to the axis in the rotated frame. Starting from $|\psi'(t_0)\rangle=|\psi(t_0)\rangle=\ket{\uparrow_z}$, we find that $|\psi'(t)\rangle$ rotates around the vector $\boldsymbol{\Omega}=(0,v,H)$. We recover that the dynamics is static in the rotating frame when $v/H\to 0$ (adiabatic limit). At non-zero velocity $v>0$, the spin rotates around $\boldsymbol{\Omega}$ and the non-adiabatic response is characterized by the angle of $\boldsymbol{\Omega}$ with the $\hat{z}$-axis. In particular, one observes a non-zero expectation value for $\av{\sigma^y (t)}$, which oscillates with frequency $\sqrt{H^2+v^2}$ and an amplitude proportional to velocity $v$. This sweep protocol is also known as Adiabatic Rapid Passage (ARP) technique, and is widely used in the magnetic resonance community to invert the population of two-level systems~\cite{Grynberg_Aspect_Fabre}. 

For a sweep described by $\theta(t)$ in Eq.~\eqref{eq:4} and $\phi=0$, the non-adibatic response corresponds to a non-zero value of $\av{\sigma^y (t)}$. As shown in Refs.~\onlinecite{polkovnikov:PNAS,polkovnikov:course} this response is closely related to the geometrical properties of the ground state via
\begin{align}
\frac{1}{2}\sin [v (t - t_0)]  \av{\sigma^y (t)} = \frac{v}{H} \mathcal{F}_{\phi=0 \theta(t)} + \mathcal{O}(v^2/H^2).
\label{curvature_sigma_y}
\end{align}
In the quasi-adiabatic regime where the occupation of the ground state remains close to one, the non-adiabatic response of the system is proportional to the Berry curvature at first order in $v/H$. The result in Eq.~\eqref{curvature_sigma_y} can be derived using a time-dependent version of the Hellmann-Feynman theorem~\cite{Hayes:chem_phys}, and proven using adiabatic time-dependent perturbation theory\cite{polkovnikov:PNAS,polkovnikov:course}. We explicitly derive it in Appendix~\ref{appendix_perturbation_theory}. 

Measuring the Berry curvature via this sweep technique was recently achieved in circuit QED experiments~\cite{Schroer:PRL,Roushan:Nature}, where the transverse spin component $\av{\sigma^y(t)}$ was measured using $\pi/2$ tomographic pulses. From a time integration over $\av{\sigma^y (t)}$, the authors estimated the Chern number $C_{\text{dyn}}$ for each value of $H_0$. In this way, they determined the location of the (Haldane) topological transition from $C = 1$ to $C = 0$ occuring at $H_0=H$ (see Eq.~\eqref{Htls_topo}). 

It may be noticed that one can reach an even more convenient expression for $C_{\text{dyn}}$ using the Heisenberg equation of motion for $\sigma^z$, which reads $\langle \dot{\sigma}^z (t) \rangle=-H \sin [\theta (t)] \langle \sigma^y (t) \rangle  $. This allows deriving a dynamical generalization of Eq.~(\ref{C_sigma_z}), which reads
\begin{equation}
  \label{eq:7}
  C = C_{\text{dyn}} + \mathcal{O}(v/H)
\end{equation}
with 
\begin{equation}
C_{\text{dyn}} = \frac{ \langle \sigma^z(t_0) \rangle - \langle \sigma^z(t_f=t_0+\pi/v) \rangle}{2} \,.
\label{C_m_sigma_z}
\end{equation}
From an experimental perspective, Eq.~(\ref{C_m_sigma_z}) shows that a measurement of the spin observables at the initial and final time is sufficient to characterize the global topology if higher order non-adiabatic corrections in $\mathcal{O}(v/H)$ are negligible. 

Using Eq.~\eqref{C_m_sigma_z} we easily recover that for a free spin $C_{\text{dyn}} = 1 + \mathcal{O}(v/H)$ when $H_0 =0$ (in fact for all $H_0 < H$), since from Eq.~(\ref{rotated_basis}) we find 
\begin{equation}
  \label{eq:6}
  \langle \sigma^z (t_f) \rangle= - \frac{ 1+\left(v/H\right)^2 \cos \left(\pi \sqrt{H^2+v^2}/v\right)}{1+\left(v/H\right)^2 }
\end{equation}
The condition $v \ll H$ is a sufficient criterion for quasi-adiabaticity in this case as the system is gapped along the complete path with gap size $H$ for $H_0 = 0$. For non-zero $H_0$ the minimal gap is $|H - H_0|$. As argued in Refs.~\onlinecite{polkovnikov:course,polkovnikov:PNAS}, the relation in Eq.~(\ref{curvature_sigma_y}) in fact also holds for gapless systems as long as the occupation of the ground state remains close to one, which also depends on the magnitude of the transition elements in addition to the energy difference. It is important to note that $C_{\text{dyn}}$ may not be equal to an integer in contrast to the integer quantity $C$. Rather, $C_{\text{dyn}}$ is an estimate for $C$ in the quasi-adiabatic regime. The recent experimental study of Refs.~\onlinecite{Schroer:PRL,Roushan:Nature} indeed report non-integer values of $C_{\text{dyn}}$ around the transition point $H_0 = H$ where the gap $\propto |H - H_0|$ closes and the quasi-adiabaticity criterion is not fulfilled. 

Before investigating the effect of dissipation, we want to emphasize that the derivation of Eqs.~(\ref{curvature_sigma_y}) and~(\ref{C_m_sigma_z}) is quite general~\cite{polkovnikov:course,polkovnikov:PNAS} and remains valid even in the presence of spin-bath coupling $\alpha> 0$ (but the gap $H$ must be replaced by its renormalized value as we show below). To access the behavior of the dynamical Chern number $C_{\text{dyn}}$ for spin-bath couplings $\alpha$ beyond the weak coupling limit $\alpha \ll 1$, we next employ the numerically Stochastic Schr\"{o}dinger Equation (SSE) approach. In the weak coupling limit, this question was addressed in Ref.~\onlinecite{vavilov} using a perturbative and Markovian Bloch-Redfield approach. This study confirmed that the dynamical Chern number $C_{\text{dyn}}$ is unaffected by the presence of the bath at weak spin-bath coupling $\alpha \ll 1$. We also note results of Ref.~\onlinecite{whitney_gefen}, where the authors studied the effect of an Ohmic bath on the Berry phase acquired by the spin along a periodic path of constant $\theta$, changing the azimuth $\phi$. They showed that in this case the environment affects the Berry phase, which is a local observable.

\subsection{Spin dynamics and topology from stochastic Schr\" odinger equation}
\label{sec:spin-dynam-topol}
As explained in the previous section, the spin dynamics $\av{\sigma^y(t)}$ (or $\av{\sigma^z(t)}$) gives access to the dynamical Chern number via Eqs.~\eqref{curvature_sigma_y}\eqref{C_m_sigma_z}. In the presence of an Ohmic bath, we calculate the spin dynamics using the numerically exact Stochastic Schr\"{o}dinger Equation (SSE) approach, which was developed in Refs.~\onlinecite{2010stoch,stochastic, Lesovik, Rabi_article, Ohmic_systems_article} (see also previous stochastic approaches to the spin-boson model~\cite{cao_sto,Stockburger_Mac,Stockburger,zhou_sto}).
This method is applicable in the regime $\alpha < 1/2$ and becomes numerically exact in the universal regime of a large bath bandwidth $\omega_c \gg H$. The SSE approach was successfully used to describe the dynamics of the Ohmic spin-boson model~\cite{2010stoch,stochastic,Ohmic_systems_article} as well as the dynamics of the Rabi model~\cite{Rabi_article}. We present a summary of the most important technical details in Appendix~\ref{appendix_SSE}. 

We calculate the dynamics of the spin expectation values $\av{\sigma^\alpha}$ for $\alpha = x,y,z$ for an external linear sweep of the polar angle $\theta(t) = v (t-t_0)$ of the magnetic field $\bfh(t)$ (see Eq.~\eqref{Htls_topo}). In Fig.~\ref{dynamics_topo}, we present results for the time-evolution of $\langle \sigma^{\alpha}(t) \rangle$ for increasing values of spin-bath coupling $\alpha$ and fixed sweep velocity $v/H=0.08$ as well as $H_0 = 0$. This choice of $v/H \ll 1$ guarantees quasi-adiabaticity in the non-dissipative case $\alpha = 0$. For a dissipation strength below $\alpha \simeq 0.15$, the $\langle \sigma^z(t) \rangle$ follows the external field and shows a complete transfer of the spin direction from $+1$ to $-1$, leading to a dynamical Chern number $C_{\text{dyn}} \simeq 1$. This confirms that global topological properties are unaffected by the presence of the environment at low dissipation. The continuous change of $\langle \sigma^z(t) \rangle$ along the path, however, becomes sharper when the coupling increases, as can be seen in the first two top panels of Fig.~(\ref{dynamics_topo}). At the same time, the amplitude of $\langle \sigma^y \rangle$ increases, especially when the field $\bfh(t)$ lies around the equator $\theta=\pi/2$. This is a clear signature of the progressive bath-induced deformation of the Berry curvature. Finally, for stronger spin-bath couplings above $\alpha \simeq 0.2$, the expectation value $\langle \sigma^z \rangle$ at the final time $t_f$ is larger than $-1$. This leads to a non-integer result for the dynamical Chern number $0<C_{\text{dyn}}<1$, which therefore ceases to be a good estimate for $C$. Since $\Delta_r \rightarrow 0$ for $\alpha \rightarrow \alpha_c = 1$ the dynamical protocol breaks down for any velocity at the localization phase transition. 

\begin{figure}[tb]
\center
\includegraphics[width=\linewidth]{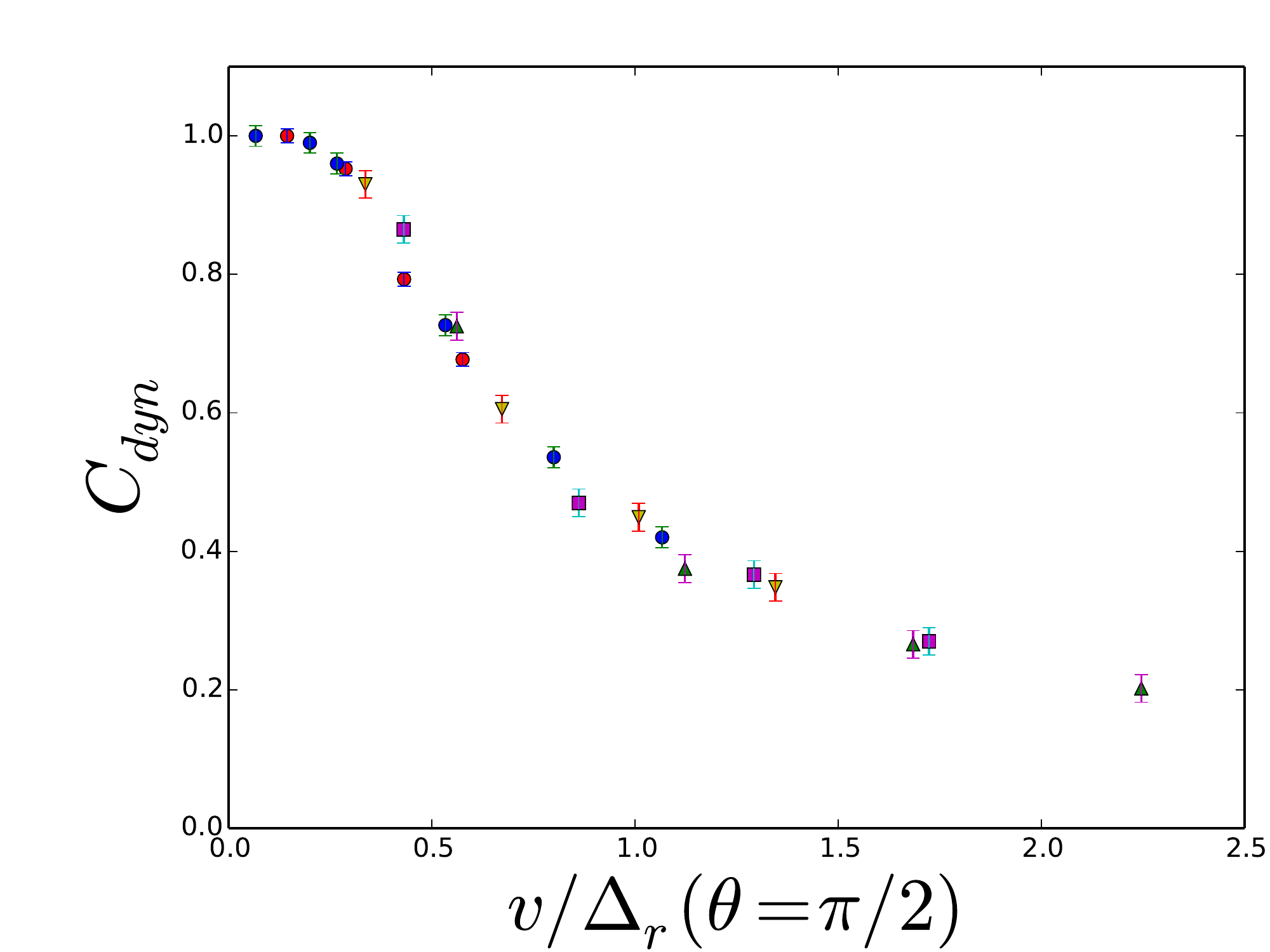}
\caption{Evolution of the dynamical Chern number $C_{\text{dyn}}$ with respect to $v/\Delta_r(\theta=\pi/2)$ for $\alpha=0.3$ (red points), $\alpha=0.36$ (blue points), $\alpha=0.38$ (yellow up pointing triangles), $\alpha=0.4$ (magenta squares) and $\alpha=0.42$ (green up-pointing triangles). $C_{\text{dyn}}$ is a good estimate for $C$ as long as the quasi-adiabaticity criterion (\ref{critere_adiabatic}) is fulfilled.}
\label{C_M_versus_v}
\end{figure}

For a given value of $\alpha$ (and $\omega_c$), one can determine the velocity at which $C_{\text{dyn}}$ stops being a good estimate for $C$ from the exact scaling of the Berry curvature at the equator $\mathcal{F}_{\phi \theta=\pi/2} = F(\alpha) \Delta/\Delta_r$ in Eq.~(\ref{F_bethe_equator}). Note that at the equator $\theta = \pi/2$ the transverse field $\Delta = H \sin \theta = H$. The dynamical Bloch spin vector $\av{\boldsymbol{\sigma}(t)}$ is able to adiabatically follow the ground state Bloch vector $\braopket{g(\theta)}{\boldsymbol{\sigma}}{g(\theta)}$ as long as the magnetic field $\bfh(t)$ evolves slowly compared to the curvature of the manifold, \emph{i.e.} as long as $\mathcal{F}_{\phi \theta=\pi/2} v/H \ll 1$ holds. Using Eq.~\eqref{F_bethe_equator}, the criterion for quasi-adiabaticity at the equator thus corresponds to
\begin{align}
v \ll \Delta_r (\theta=\pi/2)\,.
\label{critere_adiabatic}
\end{align}
where $\Delta_r=\Delta(\Delta/\omega_c)^{\alpha/(1-\alpha)}$ is the renormalized transverse field. In Fig.~\ref{topo_TLS_isole}, we illustrate the renormalized external magnetic field that the spin experiences. Note that the transverse and the longitudinal field are renormalized by the bath in very different ways. This criterion is confirmed in Fig.~\ref{C_M_versus_v}, where we show $C_{\text{dyn}}$ as a function of $v/\Delta_r$. As long as Eq. (\ref{critere_adiabatic}) is fulfilled, one can thus determine the Chern number $C$ and the bath-induced deformation of the Berry curvature $\mathcal{F}_{\phi \theta}$ via a dynamical measurement of $\langle \sigma^y \rangle $ (or $\av{\sigma^z}$). 

\begin{figure}[tb]
\center
\includegraphics[width=\linewidth]{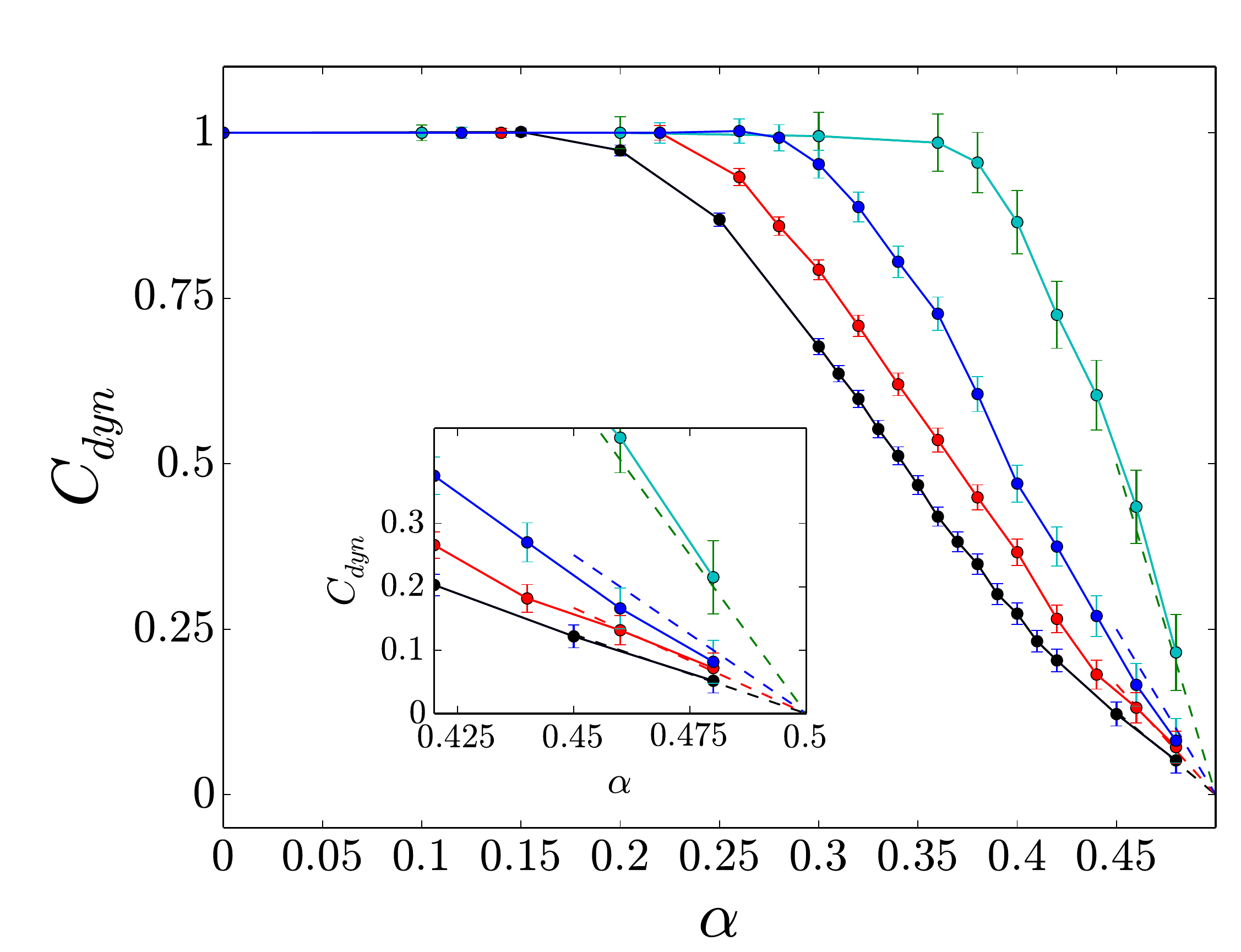} 
\includegraphics[width=\linewidth]{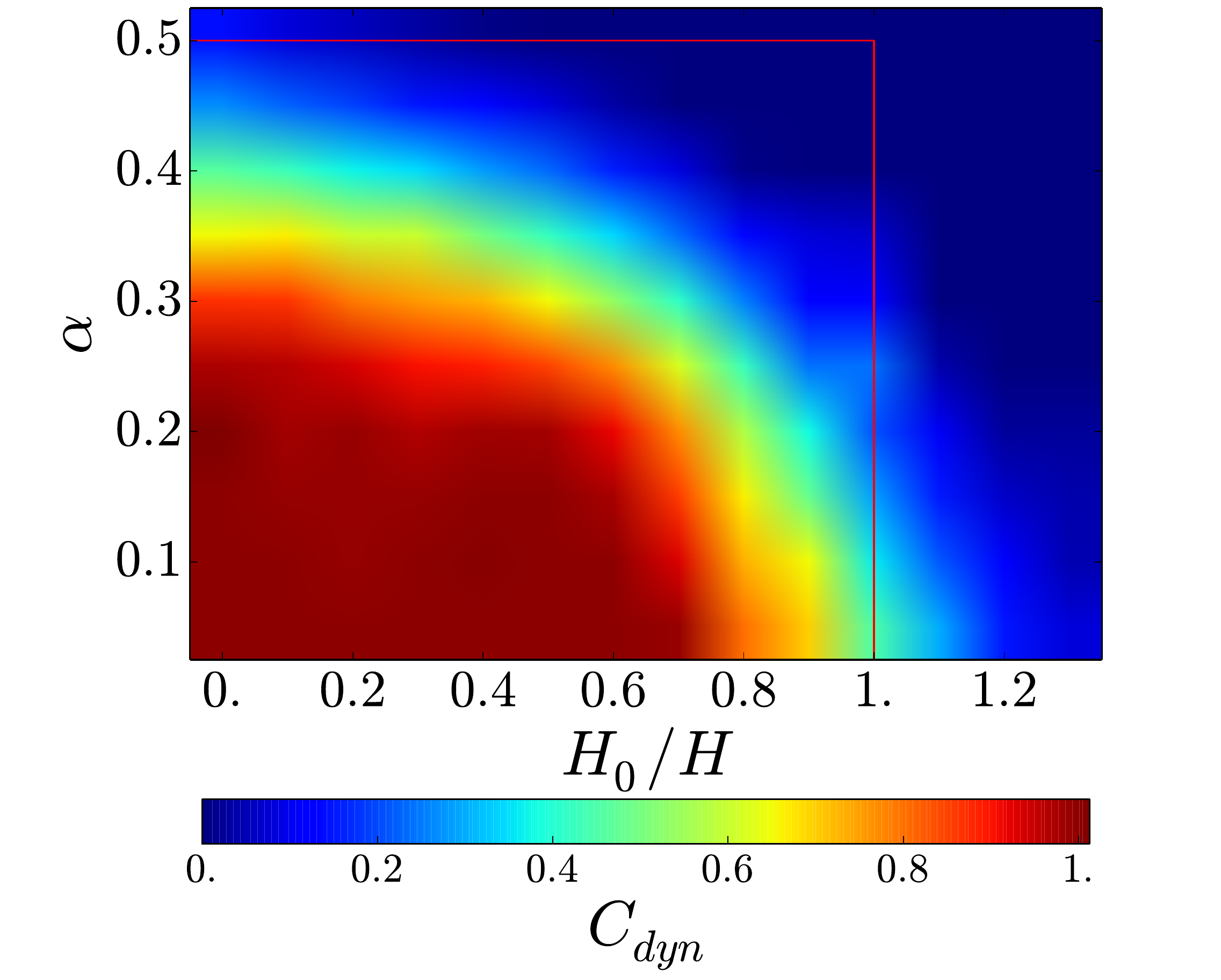} 
\caption{(Color online) (Upper panel) The main panel shows the evolution of the measured Chern number $C_{\text{dyn}}$ with respect to $\alpha$ for $v/H=0.08$ (black), $v/H=0.06$ (red), $v/H=0.04$ (blue), and $v/H=0.02$ (green). The inset zooms on the region around $\alpha=1/2$, and the dashed lines represent predictions based on the mapping with an interacting resonance level valid around $\alpha=1/2$ (see Sec. \ref{Toulouse}). We have $v/H<1$ and $v\omega_c/H^2 >1$. (Lower panel) Evolution of $C_{\text{dyn}}$ with respect to $\alpha$ and $H_0/H$ for $v/H=0.08$. The red line shows the asymptotic jump line of $C_{\text{dyn}}$ from $1$ to $0$ expected when decreasing the velocity while keeping $v\omega_c/H^2$ constant.
}
\label{chern_transition_1D}
\end{figure}

As shown in Fig.~\ref{chern_transition_1D}, this bath induced crossover from quasi-adiabatic behavior $v \ll \Delta_r (\theta=\pi/2)$ for small $\alpha$ to non-adiabatic behavior $\Delta_r \ll v \ll H$ (for all $\theta$) occurs at a coupling strength $\alpha < \alpha_c$ \emph{much smaller} than the critical value $\alpha_c = 1$. This follows from the renormalization of $\Delta \rightarrow \Delta_r$, which suppresses $\Delta_r$ to a value that decreases for larger bath bandwidth $\omega_c$. Clearly, the value of $\alpha$ where $C_{\text{dyn}}$ starts to deviate from $C = 1$ increases for decreasing velocities, corresponding to the criterion $v \ll \Delta_r (\theta=\pi/2)$ that ensures quasi-adiabatic behavior (and $C_{\text{dyn}} = C$) being fulfilled up to larger values of $\alpha$. In addition, it is interesting to note that all curves $C_{\text{dyn}}$ approach zero as $\alpha \rightarrow 1/2$. This follows from our choice of keeping $v/H \ll 1$ and $H/\omega_c \ll 1$ fixed and assume the hierarchy $H/\omega_c < v/H < 1$. This choice (of the order of limits) ensures that the dynamics is non-adiabatic at the Toulouse point, where $\Delta_r(\alpha = 1/2) = \Delta^2/\omega_c \ll v$. It enables us to quantitatively access the behavior of $C_{\text{dyn}} \rightarrow 0$ from the exact solution at the Toulouse point (see dashed lines in inset of Fig.~\ref{chern_transition_1D} and their derivation in the next Sec.~\ref{Toulouse}). Equation (\ref{critere_adiabatic}) suggests that taking the limit $H/\omega_c \to 0$ and $v/H  \to 0$ while keeping $v/H (\omega_c/H)^{\mu/(1-\mu)}$ fixed, where $\mu$ is a real number between 0 and 1, would yield a jump from $C_{\text{dyn}}=1$ to $C_{\text{dyn}}=0$ at $\alpha=\mu$.



Before discussing more quantitatively how $C_{\text{dyn}}$ behaves for larger velocities $v > \Delta_r$, let us briefly discuss the three main effects of the bath on the spin: (i) bath induced renormalization of the transverse field $\Delta = H \sin \theta \rightarrow \Delta_r = \Delta (\Delta/\omega_c)^{\alpha/(1 - \alpha)} < \Delta$ (see Fig.~\ref{topo_TLS_isole}); (ii) quasi-static bath induced bias field $\bfh_{B} \propto (0,0,2 \alpha v)$ due to the initial polarization of the bath oscillators (shifted oscillator state). The oscillators with frequencies $\omega_k < v$ are not able to follow that spin dynamics during the sweep protocol (on timescale $v^{-1}$) and thus remain in their initially polarized state. Note that we assume that the bath has intially relaxed to a shifted bath state with fully polarized spin in state $\ket{\uparrow_z}$, which is the case in our numerical protocol. (iii) resonantly induced bath bias field $\bfh_{B, \text{ind}}$ due to bath oscillator modes of frequency $\omega_k \approx v$. These modes are resonantly excited due to the sweep of the external field on a timescale $v^{-1}$. For stronger spin-bath couplings $\alpha$ these resonant bosonic modes reach large occupations causing a shift of the oscillator coordinates $x_k \propto \lambda_k (b^\dag_k + b_k)$ that opposes a further change of the spin due to the term $\sigma^z \sum_{k} \lambda_k (b^\dag_k + b_k)$ in the Hamiltonian. This behavior resembles the Faraday effect of electrodynamics with induced field $\bfh_{B, \text{ind}} (t) \propto \hat{z} \sum_{k \approx v} \lambda_k \av{b^\dag_k + b_k}(t)$ and we therefore name it a ``quantum dynamo effect''. We investigate this new effect in detail within the context of a toy model in Sec.~\ref{sec:radi-casc-phot}.

 
\subsection{Toulouse Limit}
\label{Toulouse}
To analytically access the behavior of $C_{\text{dyn}}$ as $\alpha \rightarrow 1/2$, we use that the Ohmic spin-boson Hamiltonian can be mapped onto the non-interacting resonance level (RLM) Hamiltonian at the Toulouse point $\alpha=1/2$~\cite{Toulouse,Anderson_Yuval_Hamann}. The RLM Hamiltonian describes a electronic energy level (fermionic creation operator $d^\dag$ with level energy $\epsilon_d$) that is uniformly coupled to a bath of spinless electrons described by operators $c^\dag_k$:
\begin{align}
\mathcal{H}_{RLM} =\sum_k \epsilon_k c_k^{\dagger} c_k +\epsilon_d d^{\dagger}d +V \sum_k \left( c_k^{\dagger}d+d^{\dagger} c_k \right) \,.
\label{H_toulouse}
\end{align}
Here, $\epsilon_k$ is the energy of electron $c^\dag_k$. The mapping yields a constant (fermionic) density of states over bandwidth $D$. The hybridization between level and electronic lead is denoted by $V$. The equivalence between the two models in the limit $H/\omega_c \ll 1$ can be shown by an explicit computation of the partition function~\cite{leggett:RMP,weiss}, or using bosonization~\cite{Guinea_bosonization}. One finds the following correspondence between the two models
\begin{align}
\Delta = H \sin \theta &\equiv V \sqrt{\frac{D}{4\omega_c}}\notag \\
- H_z = -H \cos \theta &\equiv \epsilon_d \,.
\label{mapping}
\end{align}
The two cutoffs are related via $D=4\omega_c/\pi$~\cite{Hur}. Note that the mapping becomes formally exact in equilibrium in the scaling limit of infinite bandwidth $\omega_c \rightarrow \infty$ (keeping $\Delta_r(\alpha = 1/2) = \Delta^2/\omega_c$ fixed~\cite{PhysRevB.88.165133}) and in the dynamics at times larger than $t > \omega_c^{-1}$.~\cite{leggett1987dynamics}

The Toulouse RLM Hamiltonian in Eq.~\eqref{H_toulouse} is non-interacting. We can therefore exactly solve the dynamics of, e.g., the level occupation $\av{d^\dag d}(t)$, for the sweep protocol $\theta(t)=v(t-t_0)$ using the Keldysh technique~\cite{antti_pekka_wingreen_meir}. The dot occupation can be related to the spin expectation value $\langle \sigma^z(t) \rangle$. From the equations of motion, we find
\begin{align}
\partial_t \langle d^{\dagger}d \rangle= iV \sum_k \left[ \langle c_k^{\dagger} d \rangle-\langle d^{\dagger} c_k  \rangle \right] \,.
\label{EOM_toulouse}
\end{align}
The computation of the right hand side of Eq.~(\ref{EOM_toulouse}) can be performed exactly, and we obtain an expression similar to the one obtained for the lead currents in Ref.~\onlinecite{antti_pekka_wingreen_meir} (see Eq.~(42) of this reference),
\begin{align}
\partial_t \langle d^{\dagger}d \rangle& = -\Gamma(t) \langle d^{\dagger}d \rangle -\int d\epsilon f(\epsilon)/\pi  \\
& \quad \times \int_{t_0}^t dt_1 \Gamma(t_1,t) \Im m \left\{ e^{-i \epsilon (t_1-t)} G^r (t,t_1) \right\} \nonumber \,.
\label{EOM_toulouse_2}
\end{align}
Here, $f$ denotes the Fermi distribution (at $T=0$ here) and we have defined 
\begin{align}
\Gamma(t,s)&=\frac{2\pi}{D} V(t) V(s),\\
G^r(t,s)&=-i \theta(t-s) \exp\left[-i \int_{s}^t du \epsilon_d (u) \right] \nonumber \\
& \quad \times \exp\left[-\frac{1}{2} \int_{s}^t du \Gamma (u) \right]\,.
\label{coeffs_EOM_toulouse}
\end{align}
We also write $\Gamma(t)=\Gamma(t,t)$. We solve the Keldysh equations~(\ref{EOM_toulouse_2}) in the scaling regime, defined by taking $\omega_c$ (or $D$) to infinity, while keeping $H^2 t /\omega_c$ fixed for all times (considering dynamics on timescales $\Delta_r^{-1}$). 

We confirm numerically that $\langle d^{\dagger}d  \rangle(t_f=t_0+\pi/v) \simeq 1$ for $\alpha=1/2$ in our regime of $\Delta_r < v < H$. This immediately implyies that $C_{\text{dyn}}=0$ at $\alpha=1/2$. The resonant level remains occupied during the dynamics, even though its final energy is above the Fermi level. This rather counter-intuitive result applies to our particular (experimentally motivated) regime where $v \gg H^2/\omega_c$, corresponding to the bandwidth $\omega_c$ being the largest energy scale and the velocity $v$ is small compared to $H$ but not compared to $H^2/\omega_c$. In this case the final time of the protocol $t_f$ is always much smaller than the typical time-scale of the Rabi dynamics at the Toulouse point $\Delta_r^{-1} = \omega_c/\Delta^2$ (see first term in Eq.~\eqref{EOM_toulouse_2}). Note that at the ``equator'' ($\theta = \pi/2$), the change in the dot occupation number is maximal, since the level is resonant with the Fermi energy, the relation $\Delta = H \sin \theta$ implies $\Delta = H$. Here, it is important to stress that the non-interacting resonant fermionic level model yields $C_{\text{dyn}} = 1$ in the adiabatic regime $v \ll \Delta_r = \Delta^2/\omega_c$, again confirming the quasi-adiabaticity criterion in Eq.~\eqref{critere_adiabatic}. 

\subsection{Scaling of $C_{\text{dyn}}$ close to $\alpha = 1/2$}
\label{sec:emerg-fermi-liqu}
To explore the vicinity of the Toulouse point with $\alpha < 1/2$, we define the dimensionless variable $u = \frac12 - \alpha \ll 1$. Non-zero $u$ result in an additional interaction term $\mathcal{H}_u$ in the RLM (see Eq.~\eqref{H_toulouse}) of the form~\cite{Guinea_bosonization}
\begin{align}
\mathcal{H}_u = U\sum_{k,k'} \left( c^{\dagger}_k c_{k'}-\frac{1}{2}\right) \left( d^{\dagger} d-\frac{1}{2}\right)
\label{additional_term}
\end{align}
with $U=\pi \left(1-\sqrt{2\alpha} \right) = \pi u + \mathcal{O}(u^2)$. The inclusion of this interaction term in the Keldysh formalism, which describes an interaction between the electron on the level and at the first site of the lead, hinders the closure of the equations of motion for the Green's functions. In Appendix~\ref{appendix_keldysh}, we treat this interaction term in a basic mean-field approximation that enables us to numerically compute the time evolution of $\langle d^{\dagger}d \rangle(t)$ for $\alpha < 1/2$. Interestingly, we numerically find a linear behaviour 
\begin{equation}
  \label{eq:8}
  C_{\text{dyn}}(u)= a \frac{H}{v} u
\end{equation}
with $a$ constant. The dynamical Chern variable thus vanishes at $\alpha = 1/2$ and increases linearly in $\alpha$ for $\alpha<1/2$ with a slope that diverges as $v/H \rightarrow 0$ (note that we always demand $v/H > \Delta^2/\omega_c$). In Fig.~\ref{chern_transition_1D} we show that this scaling prediction is in very good agreement with the numerical result of $C_{\text{dyn}}$ obtained from the spin-boson model.

This linear dependency may be interpreted in the analogy to the Fermi-liquid behaviour of the Ohmic spin-boson model, or its Kondo analogue~\cite{Nozieres}. In this description, the local susceptibility $\chi=-\partial_{H_z} \langle \sigma^z \rangle $ is known to be constant with respect to $H_z$, the $z$-component of the vector $\boldsymbol{H}$ (see Eq.~\eqref{Htls_topo}). A deviation $u>0$ from the point $\alpha=1/2$ may thus be seen as a shift of the electronic level energy by a factor proportional to $u$, or equivalently a shift of $H_z$ in the spin-boson description. This argument would confirm then a linear dependence of $\langle \sigma^z (t_f=t_0+\pi/v) \rangle$ with respect to $u$, and thus the scaling $C_{\text{dyn}}(u) \propto u$ that we find in Eq.~\eqref{eq:8}. The fact that the slope scales as $v/H$ and the curve thus becomes very steep for small $v/H \ll 1$, bears similarities with the dependence of $\chi \propto 1/T_K$ in the anisotropic Kondo model, where $T_K \sim \Delta_r$ is the Kondo temperature, which diverges at the antiferromagnetic-ferromagnetic quantum phase transition $\alpha_c=1$.


\section{Radiative cascade of photons: Quantum Dynamo Effect}
\label{sec:radi-casc-phot}
In the previous section we have found a bath induced crossover from quasi-adiabatic behavior at small $\alpha$, where $v \ll \Delta \approx \Delta_r$, to non-adiabatic behavior $v \gg \Delta_r$ as the spin-bath coupling $\alpha$ (or alternatively $\omega_c$) is increased. We pointed out three effects that the bath has on the spin: (i) renormalization of $\Delta$ to $\Delta_r < \Delta$, leading to a reduction of the minimal gap at the equator; (ii) static bath bias field $\bfh_{B} \propto \alpha v \hat{z}$ and (iii) resonantly induced bath bias field $\bfh_{B, \text{ind}} \propto \hat{z} \sum_{k \approx v} \lambda_k \av{b^\dag_k + b_k}$. Both effects (ii) and (iii) tend to increase the magnetic field along the $\hat{z}$ axis.
While the first two effects are well established and have been studied in detail previously~\cite{weiss,stochastic}, we newly identify the third effect of a resonantly induced bath bias field here. In the regime of small velocities $v \ll H$ considered here, the resonantly induced bath bias field (due to effect (iii)) turns out to be much larger than the quasi-static bath bias field (due to effect (ii)).

In Fig.~\ref{C_M_versus_v}, we observe that $C_{\text{dyn}} \rightarrow 0$ as $\alpha \rightarrow 1/2$. This corresponds to a situation where the spin does not follows the external magnetic field sweep at all and $\av{\sigma^z(t_f)} \approx \av{\sigma^z(t_0)}$ remains close to its initial value. While $v \ll \Delta_r$ explains the breakdown of (quasi-)adiabaticity, which is a necessary condition that $C_{\text{dyn}} = C$, the fact that $C_{\text{dyn}} \rightarrow 0$ additionally signals that the effective magnetic field that the spin experiences at the end of the sweep $t=t_f$ is still along the positive $\hat{z}$ direction. 

As the quasi-static bath induced bias field $\bfh_{B} \propto \alpha v$ cannot compensate the external field $H_z(t_f) = - H$ in the regime $v/H$ we consider, this field compensation is clearly due to another effect. As we demonstrate below, it is due to resonantly excited bath modes caused by driving the spin at velocity $v$. This not only prevents the complete system to remain in its instantaneous ground state and leads to pronounced spin-bath entanglement, which we discuss in detail in the next Sec.~\ref{sec:entangl-entr-effect}. The bath excitations act on the spin as an effective magnetic field along the $\hat{z}$-direction
\begin{equation}
  \label{eq:9}
  \bfh_{B, \text{ind}} \propto \hat{z} \sum_{k \approx v} \lambda_k \av{b^\dag_k + b_k} \,.
\end{equation}
Due to the resonance condition, a large number of bosons is created and this strongly affects the spin polarization polarization. At sufficiently large $\alpha$, it can compensate the external field $-H \hat{z}$ present at the end of the sweep such that $\bfh_{B, \text{ind},z} > H$ and the total field is along the positive $\hat{z}$ direction, preventing the spin to flip. Note that this corresponds to the spin path on the right of Fig.~\ref{topo_TLS_isole}.



\begin{figure}[tb]
\center
\includegraphics[width=\linewidth]{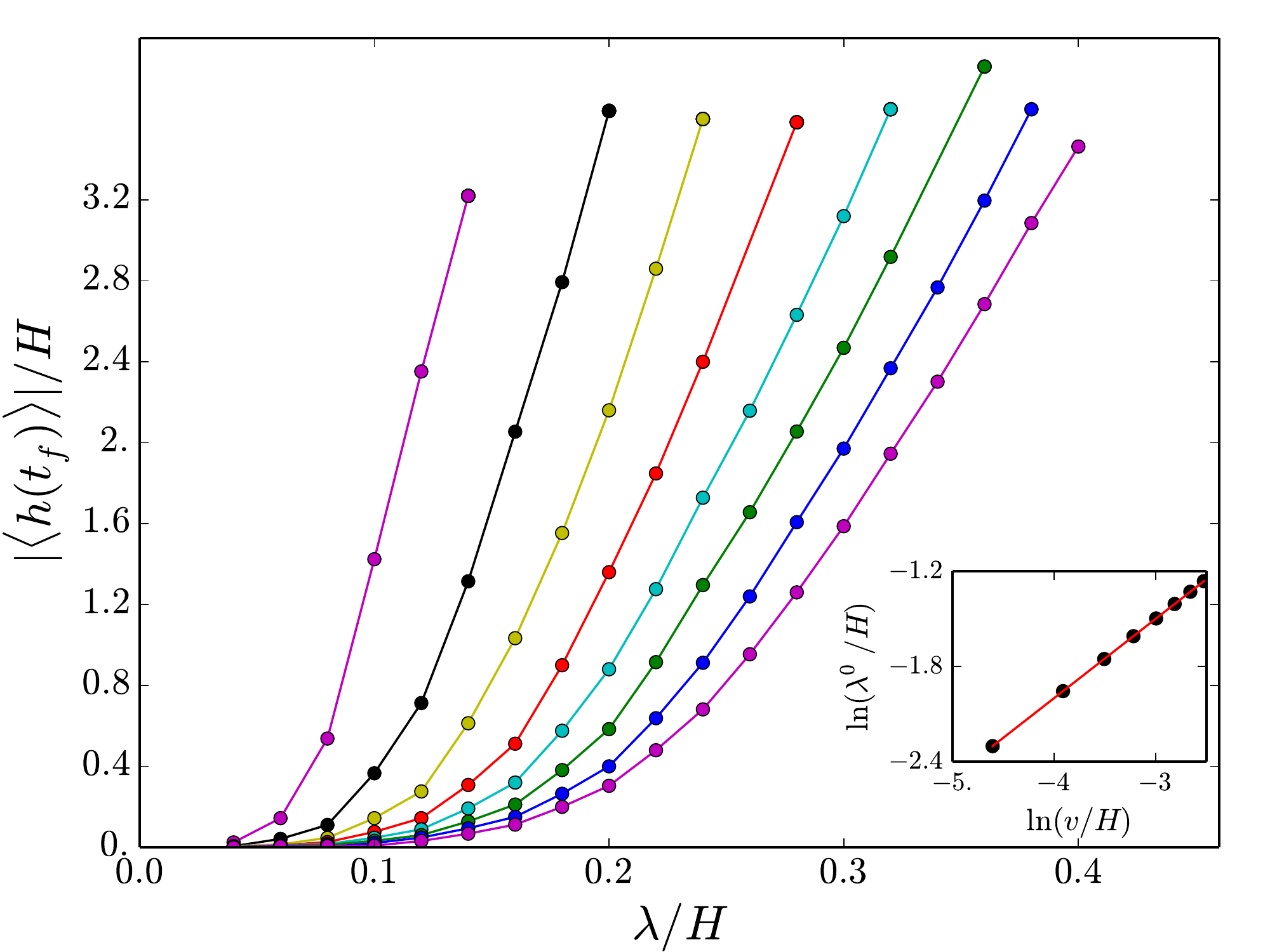}    
\caption{Effective field $|\langle h \rangle|/H$ felt by the spin at the end of the dynamical protocol as a function of $\lambda$, for $v/H=0.01$ (magenta), $v/H=0.02$ (black), $v/H=0.03$ (yellow), $v/H=0.04$ (red), $v/H=0.05$ (cyan), $v/H=0.06$ (green), $v/H=0.07$ (blue), $v/H=0.08$ (magenta). Inset: Fit of the value of the coupling $\lambda^0$, for which $|\langle h \rangle|/H$=1, giving the scaling $\lambda^0/H=  (v/H)^{1/2}$.}
\label{Results_toy_model}
\end{figure}

To illustrate this effect, we numerically study a simpler single-mode toy model
\begin{align}
  \mathcal{H}_{\text{single-mode}} &= \frac{H}{2} \cos(vt) \sigma^z+  \frac{H}{2} \sin(vt) \sigma^x \nonumber \\
  & \qquad + \frac{\lambda}{2} \sigma^z(b+b^{\dagger}) + v b^{\dagger}b \,.
\label{Toy_model}
\end{align}
This model only considers the effect of a single mode of frequency $v$. The evolution of the Bloch vector $\boldsymbol{\sigma}(t)=\langle \boldsymbol{\sigma} (t)\rangle$ is given by
\begin{align}
\partial_t \boldsymbol{\sigma} =\boldsymbol{H} \times  \boldsymbol{\sigma},
\label{Bloch_equation}
\end{align}
where $\boldsymbol{H}=\boldsymbol{H}+ h_{\text{ind}} \sigma_z(t)$ and $h_{\text{ind}}(t) = \lambda \av{b+b^{\dagger}}(t)$. The time evolution of $h_{\text{ind}}(t)$ is given by
\begin{align}
\frac{1}{v^2} \partial_{t}^2 h_{\text{ind}} + h_{\text{ind}} =-\frac{\lambda^2}{v} \sigma_z(t) \,,
\label{Evolution_h}
\end{align}
This equation describes a harmonic oscillator driven by an inhomogeneity $\propto \sigma_z(t)$. The strength of the drive is proportional to $\lambda^2/v$, which can be solved straightforwardly. In Fig.~\ref{Results_toy_model}, we show the absolute value of the effective field $|\av{h}(t_f)|$ that the spin experiences at the end of the sweep. The field increases sharply as a function of spin-mode coupling $\lambda$ at a particular value of $\lambda$ that depends on the sweep velocity $v$. This shows that a particular coupling strength $\lambda$ is necessary to trigger the resonant excitation of the single mode and reach substantial values of the effective field $h(t_f)$.
At a fixed value of $\lambda$, the field $|h_{\text{ind}}(t_f)|$ decreases with velocity $v$. We numerically extract the general behaviour $|\langle h_{\text{ind}} (t_f) \rangle| \propto \lambda^2/v$. Setting $\lambda=\sqrt{2\alpha v H}$ in agreement with an Ohmic spectral function at low frequencies, we recover that
\begin{equation}
  \label{eq:10}
  |\langle h_{\text{ind}} (t_f) \rangle| \propto 2\alpha H > H 
\end{equation}
for sufficiently strong $\alpha$. This demonstrates that the field $h_{\text{ind}}(t_f)$ can fully compensate the external magnetic field for sufficiently large $\alpha$ (for our choice of $\lambda$, we find indeed $\alpha=1/2$ as the critical coupling strength). It is important to recall that in the spin-boson model the corresponding phenomenon is a many-body effect. The toy model, however, provides an intuitive picture of how a radiative cascade of bosons leads to a bath induced magnetic field $h_{B, \text{ind}}$ that is large enough to compensate the external field. Due to the similarity with the electromagnetic Faraday effect, we name this a ``quantum dynamo effect''.

\section{Entanglement entropy and effective thermodynamics}
\label{sec:entangl-entr-effect}
Let us finally study the entanglement between spin and bath and provide an effective thermodynamical description of the final state of the system at $t_f$ after the sweep. This will allow us to identify the dissipation strength $\alpha_0$, where the crossover from quasi-adiabatic to non-adiabatic behavior occurs (see Fig.~\ref{chern_transition_1D}), as a region of maximal spin-bath entanglement and population inversion (corresponding to effectively negative temperatures). 

The entanglement entropy describes the amount of entanglement that is present between the spin and the bath modes~\cite{entanglement_entropy}. It is defined as 
\begin{equation}
  \label{eq:11}
  \mathcal{E}=-\text{Tr}\left[ \rho_S \log_2 \rho_S\right] \in [0,1] \,,
\end{equation}
where $\rho_S$ is the spin-reduced density matrix $\rho_s = \text{Tr}_B(\rho)$. For a pure state, spin and bath density matrices are factorized yielding $\mathcal{E}=0$. In contrast, the case $\mathcal{E}=1$ corresponds to a maximally entangled spin-bath state. 

\begin{figure}[tb]
\center
\includegraphics[width=.8\linewidth]{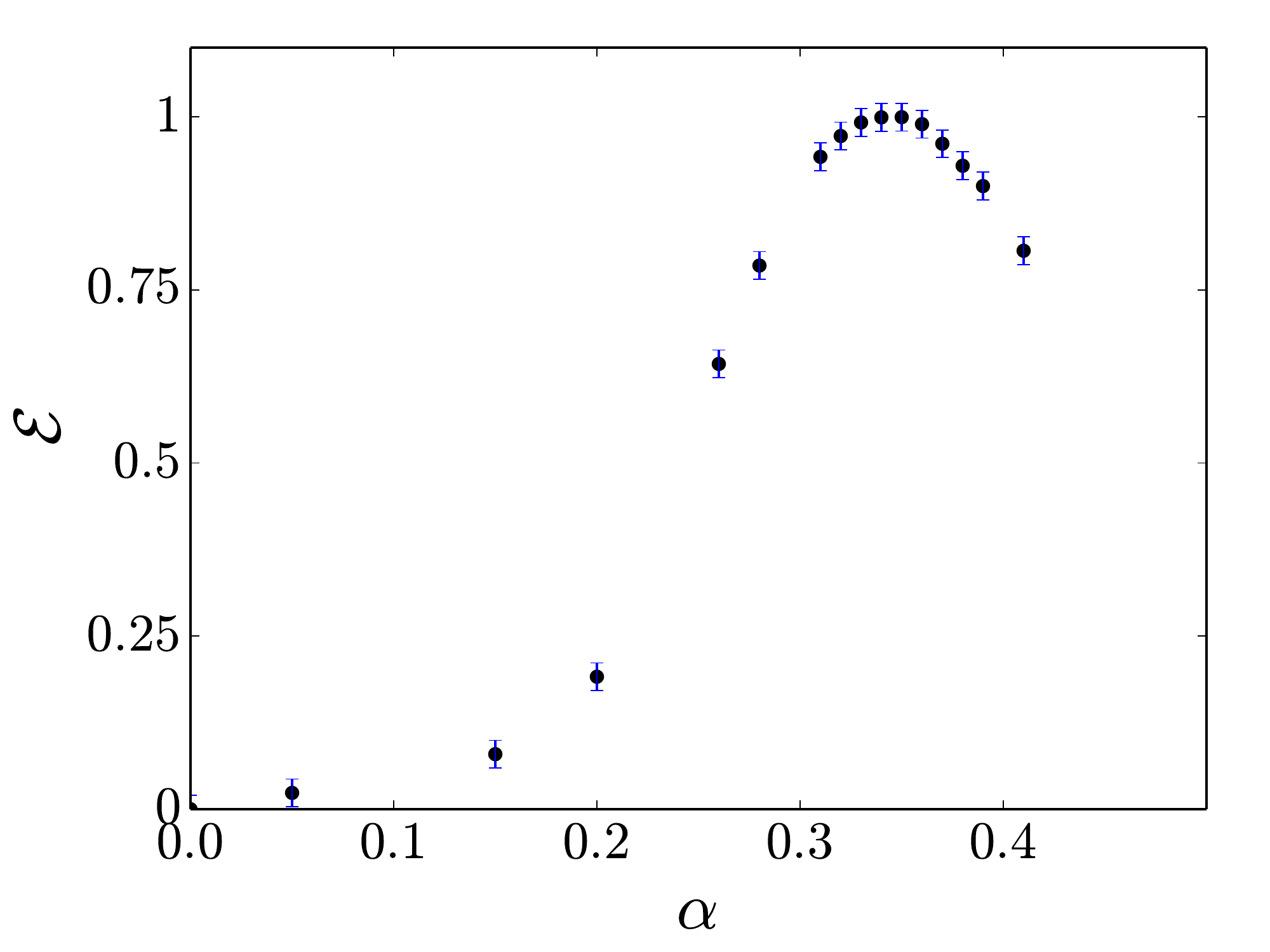}
\includegraphics[width=.8\linewidth]{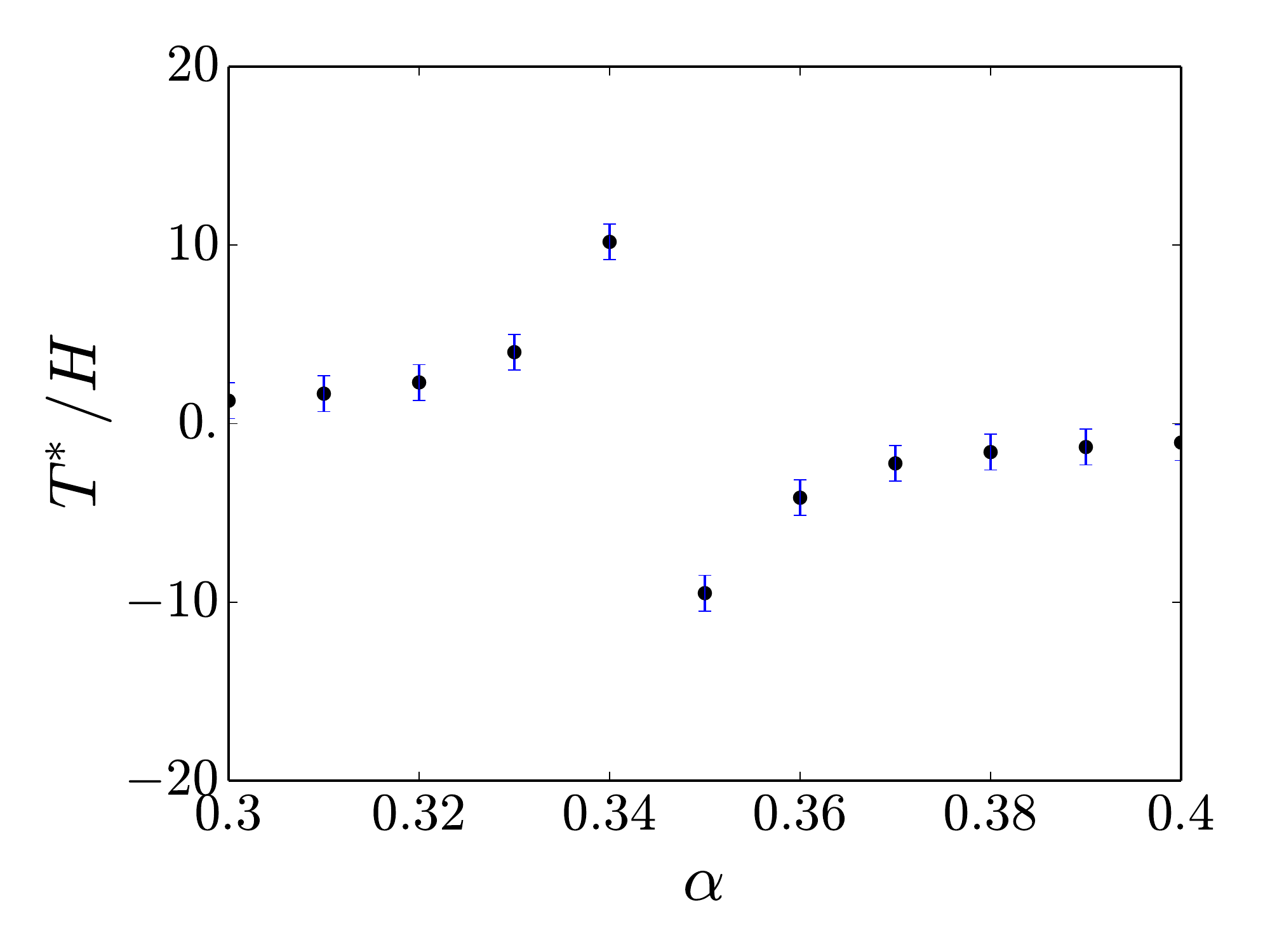}
\caption{(Upper panel) Evolution of the final entanglement entropy (Left panel), and effective temperature as a function of $\alpha$ for fixed $v/H=0.08$ and $H/\omega_c=0.01$. (Lower panel) Effective temperature $T^*/H$ as a function of $\alpha$. The temperature $T^*$ becomes negative when $\langle \sigma^z (t_f) \rangle=0$ and the entanglement entropy $\mathcal{E}$ is maximal.}
\label{criticality}
\end{figure}

In the upper panel of Fig.~(\ref{criticality}), we show entanglement entropy in the final state after the sweep as a function of $\alpha$ for fixed $v/H=0.08$ and $H/\omega_c = 0.01$. We observe that the entanglement between spin and bath increases with spin-bath coupling $\alpha$ and reaches its maximum $\mathcal{E}=1$ at the characteristic coupling $\alpha_0$, where the crossover into non-adiabatic spin dynamics occurs. For larger values of $\alpha$, the entanglement $\mathcal{E}$ decreases again as the system evolves towards a factorized spin-bath state of the form $\ket{\uparrow_z}\otimes \ket{\chi_{\uparrow}}$. It is important to note that $\mathcal{E}$ evolves smoothly with $\alpha$, in contrast to the well-known discontinuity that occurs in the Ohmic spin-boson model in equilibrium at the localization phase transition at $\alpha_c = 1$ in the absence of a field in the $\hat{z}$-direction ($H_z = 0$). We note that recently, the entanglement entropy was successfully measured experimentally in a circuit QED setup~\cite{measure_entanglement_entropy}. 

One obtains an effective thermodynamic description of the final spin reduced density matrix $\rho_s(t_f)$ by regarding it as the density matrix of an isolated spin in thermal equilibrium with an effective Hamiltonian $\mathcal{H}^*$ and an effective temperature $T^*$. Both $\mathcal{H}^*$ and $T^*$ are uniquely determined by the spin expectation values $\av{\sigma^\alpha}$. Following Ref.~\onlinecite{williams_hur_jordan}, we interpret $\rho_S(t_f)$ as the density matrix of a factorizable state $\rho=\rho_S \otimes \rho_B$ with the spin being in the thermal equilibrium at temperature $T^*$ and with effective Hamiltonian $\mathcal{H}^{*}$. Although spin and bath are in fact highly entangled, such an effective description can nevertheless be useful in order to characterize the general behaviour of the system in the vicinity of $\alpha_0$. In particular, the formal identification allows to compute the effective temperature $T^*$, whose evolution with respect to $\alpha$ is shown in Fig.~\ref{criticality}. For $\alpha> \alpha_0$, we observe negative temperatures $T^* < 0$, which indicates a population inversion in the final state of the sweep. As pointed out previously, for $\alpha > \alpha_0$ the spin no longer follows the external magnetic field (since $|\bfh_{B,\text{ind}}| > H_z$) and the final spin orientation is against the external field. We note that negative effective temperatures have been realized in localized spin systems~\cite{negative_spin_1,negative_spin_2}, and were recently measured experimentally for motional degrees of freedom in a cold atom setup~\cite{negative_temp}. Interestingly, we note that $T^*$ diverges at $\alpha_0$ with leading behaviour $T^*/H \propto 1/(\alpha-\alpha_0)^{-1}$, independently of $v/H$. This scaling behaviour may be of interest for future experimental studies.

\section{Conclusion and experimental perspectives}
\label{experimental_realizations}
Experimentally controlling the form of a quantum dissipative environment is a challenging task, but various experimental platforms have been devised in this goal. An Ohmic bosonic bath can be engineered through a long transmission line~\cite{Cedraschi:PRL,Cedraschi:Annals_of_physics}. Another setup of interest is a one-dimensional Luttinger liquid~\cite{InesSaleur,KLH2}, where a dissipative quantum phase transition were recently observed~\cite{QPT_LL,QPT_LL_1}. Cold atom setups~\cite{recati_fedichev,orth_stanic_lehur,Carlos_scientific_reports} are also viable candidates to simulate an Ohmic bath coupled to a two-level system. These platforms have already provided reliable measurements of topological and geometrical characteristics of Bloch bands~\cite{Experimental_Berry_1,Experimental_Berry_2}. It was also proposed to unravel these many body effects in a non-equilibrium protocol, by studying the transmission spectrum of microwave light in a related circuit QED scheme~\cite{photonic_kondo,hur2015many}. These circuit QED architectures appear promising as in recent experiments~\cite{Haeberlein_arxiv,Forn_diaz} values of the dissipation strength $\alpha > 0.01$ were reported. A value of as large as $\alpha \approx 0.5$ was recently reached in Ref.~\onlinecite{Forn_diaz}. An advantage of these circuit QED setups is that the quasi-adiabatic sweep protocol that we discuss in this article has already been successfully implemented~\cite{Schroer:PRL,Roushan:Nature}. Finally, solid-states architectures have also led to a measurement of the Berry phase using an interferometry protocol~\cite{Observation_Berry}.

To conclude, in this article we have investigated the topology of a spin-1/2 in contact with an Ohmic bath. We have shown that within the delocalized phase, where $\alpha < \alpha_c = 1$, dissipation modifies the geometry of the spin ground state manifold on the Bloch sphere only locally. The global topology captured by the Chern number $C$ remains unchanged by the coupling to the bath until the localization quantum phase transition at $\alpha_c = 1$. We provided a geometric interpretation discussion of this well-known transition relating the Berry curvature $\mathcal{F}_{\phi \theta}$ to the spin susceptibility that is known from exact Bethe Ansatz results. We found that the bath gradually deforms the spin ground state manifold. At $\alpha_c=1$, the Berry curvature becomes infinite and this singularity signals the dissipative topological transition. We then investigated this geometrical change of the Berry curvature withing a recently proposed dynamical sweep protocol~\cite{polkovnikov:PNAS} that relates the quasi-adiabatic response of a slowly driven spin to $\mathcal{F}_{\phi \theta}$. As long as one ensures (quasi-)adiabaticity the dynamically measured Chern number $C_{\text{dyn}}$ is a good estimate of $C$. Using numerically exact results obtained from the stochastic Schr\"{o}dinger equation (SSE) technique, we were able to identify a bath induced crossover from quasi-adiabatic to non-adiabatic behavior that occurs at fixed velocity $v/H \ll 1$ at dissipation strength $\alpha$ much smaller than the critical value $\alpha_c = 1$. Our results show that the dynamic protocol requires much smaller velocities $v \ll \Delta_r$ in the dissipative system as compared to the case of a free spin $v \ll H$. For fixed velocities in the regime $H/\omega_c < v < H$, we showed that the dynamically measured Chern number vanishes at the Toulouse point $\alpha =1/2$, and we derived analytically the scaling of $C_{\text{dyn}}$ close to $\alpha = 1/2$. We provided an intuitive physical explanation of this effect as a consquency of a resonantly induced bath bias field $\bfh_{B, \text{ind}}$ due to the resonant excitation of modes with frequencies close to the sweep velocity $\omega_k \approx v$. Using a simplified single-mode toy model, we were able to show that these bath excitations crucially affect the spin dynamics in analogy to the Faraday effect of induction in electromagnetism. We thus named this phenomenon the ``quantum dynamo effect''. 

Finally, we note that it would also be interesting to investigate the effect of other kinds of dissipative environments on the spin topology. Examples are a bosonic bath with sub-Ohmic spectral density, $J(\omega)\propto \omega^s \omega_c^{1-s}$ at low frequency with $0<s<1$, which is known to trigger a continuous quantum phase transition in equilibrium~\cite{weiss, kehrein}. The main task would be to characterize the evolution of $\langle \sigma^z \rangle$ with respect to $\theta$. From Eq. (\ref{Berry_connection_sigma_z}) and the study of critical exponents in Ref.~\onlinecite{vojta}, one expects a divergence of the Berry curvature at the equator at the transition $\mathcal{F}_{\phi \theta=\pi/2}\propto (\alpha_c-\alpha)^{-\gamma}$, where $\gamma=1+\mathcal{O}(s)$. One could also use a variational approach similar to the one used in this article or implemented in Ref.~\onlinecite{plenio_polaron} in order to reach an estimate of $\mathcal{F}_{\phi \theta}$ for all values of $\theta$. In a broader context, we note a recent observation~\cite{DQPT} of a dynamical topological transition after a quench in a system of ultracold atoms in an optical lattice, or studies focusing on the dissipative preparation of topological states in cold atomic lattice systems~\cite{Diehl_1,Diehl_2}. Beyond its theoretical interest, the experimental evidence of the bath induced adiabatic to non-adiabatic dynamical crossover and the associated quantum dynamo effect studied here, seems an accessible and exciting opportunity in state-of-the-art experimental platforms.

 \section*{Acknowledgments}
 
 We thank D. Chang, B. Dou\c{c}ot, J. Esteve, J. Gabelli, I. Garate, L. Herviou, A. Jordan, J. Keeling, P. Lecheminant, C. Neill, A. Petrescu, K. Plekhanov, Z. Ristivojevic, P. Roushan, L. Sanchez-Palencia, J. Stockburger, M. Schir\'{o} and W. Zwerger. We acknowledge financial support from the PALM Labex, Paris-Saclay, Grant No. ANR-10-LABX-0039 and by the German Science Foundation (DFG) FOR2414. We also acknowledge discussions at CIFAR meetings, at the ``Qlight" conference in Crete and at the workshop ``Simulating Quantum Processes and Devices" in Bad Honnef.

\appendix
\section{Chern number and relation to the equilibrium properties of $\langle \sigma^z \rangle$}
\label{appendix_index}

A more rigorous characterization of the relative integer $n$ introduced in the main text can be done with homotopy theory, as exposed in Ref. \onlinecite{Mermin_topo}, by considering a group $G$ of transformations between possible equilibrium Bloch vectors. In our case, a suitable group $G$ would correspond to the two-dimensional rotation group $SO(2)$, which is a continuous group with a well-defined topology. One can show in particular that to each loop in $SO(2)$, is associated a relative integer $n$\footnote{Formally, the fundamental group of $SO(2)$ is the additive group of integers $\mathcal{Z}$}. The continuous path parametrized by $\theta$ from $\theta=0$ to $\theta=2 \pi$ defines a loop in $SO(2)$, and the cases $H_0/H<1$ or $H_0/H>1$ correspond to a different value of $n$. This integer $n$ corresponds to the Chern number characterizing the spin 1/2 system.

One can use the Poincar\'{e}-Hopf theorem to show that the Chern number is equal to the degree (introduced below) of the mapping $(\theta,\phi)\to \boldsymbol{h}/|\boldsymbol{h}|$, as used for example in Ref. \onlinecite{fuchs_simon_brouwer}. The degree $deg$ of a smooth map $f: M\to N$ between two connected, oriented and closed $n$-dimensional manifolds $M$ and $N$, is an integer defined by\cite{degree1,degree2}
\begin{align}
deg=\sum_{x \in f^{-1}(y)} \textrm{sign} \det\left(\mathcal{J}\right),
\label{degree}
\end{align}
where $\mathcal{J}$ is the Jacobian matrix of $f$ and $y\in N$ is a regular point with a finite number of preimages. $deg$ is an integer which do not depend on the point $y$. In our precise case, we work with $2$-dimensional manifolds, that fulfill the requirements of the above definition. When $H_0<H$, any regular point $y$ in $N$ has only one pre-image. The sum in Eq. (\ref{degree}) reduces then to one term and we have in general $deg=\pm 1$. When $H_0>H$, the situation is different as there are always two preimages of any regular point in $N$. A computation of the Jacobian for a particular choice of $y$ shows that these two terms compensate and one gets $deg=0$. We recover then Eq. (\ref{C_sigma_z}), when we consider the limit $y\to (0,0,1)$ .

\section{Shifted oscillators approach-single polaron study}
\label{Appendix_polaron}
Recent variational approaches, such as the polaron expansion\cite{bera:PRB}, allow to determine approximatively the quantities $p$ and $q$, and states $|\chi_{\sigma}\rangle$. At weak dissipation, one may indeed approximate the ground state $|g \rangle$ in the ``single polaron" picture\cite{bera:PRB} (see also related Refs. \onlinecite{Silbey_Harris,leggett:RMP,weiss,Hur}) where one assumes that bath states $|\chi_{\sigma}\rangle$ correspond to multi-mode coherent states $|\chi_{\sigma}\rangle=\exp\left[\sum_k f_k^{\sigma} (b_k -b_k^{\dagger}) \right]|0\rangle$. This ``single polaron" picture is also often called ``shifted oscillators" picture, as a coherent state corresponds to the ground state of an harmonic oscillator whose equilibrium position has been shifted. Here, $|0\rangle$ denotes the vacuum with all the oscillators at equilibrium and $f_k^{\sigma}$ corresponds to the value with which the oscillator $k$ is shifted for the state $|\chi_{\sigma}\rangle$. The set of real numbers $\{f_k^{\uparrow }\}$ and $\{f_k^{\downarrow }\}$, as well as $p$ and $q$, are then determined by minimizing the mean energy $E=\langle g | \mathcal{H} |g\rangle$ of the system. For simplicity, we work at $\phi=0$ and we reach
\begin{widetext}
\begin{align}
E=\frac{1}{p^2+q^2}\left[\frac{H}{2} \cos \theta (p^2-q^2)+H p q\sin \theta  e^{-\sum_k \frac{\left(f_k^{\uparrow}-f_k^{\downarrow}\right)^2}{2}}+ \sum_k \lambda_k \left(p^2 f_k^{\uparrow}-q^2 f_k^{\downarrow}\right)+\sum_k \omega_k \left(p^2 (f_k^{\uparrow})^2+ q^2 (f_k^{\downarrow})^2 \right)  \right].
\label{Energy}
\end{align}
Minimizing $E$ with respect to $f_k^{\uparrow}$ and $f_k^{\downarrow}$ gives for all $k$,
\begin{align}
p^2 \lambda_k  +2p^2 f_k^{\uparrow}\omega_k -p q H\delta \sin \theta(f_k^{\uparrow}-f_k^{\downarrow})=0 \label{minimization_f_k_g_k_0}\\
-q^2 \lambda_k  +2q^2 f_k^{\downarrow}\omega_k +p q H \delta \sin \theta(f_k^{\uparrow}-f_k^{\downarrow})=0 
\label{minimization_f_k_g_k}
\end{align}
where $\delta=e^{-\sum_k \frac{\left(f_k^{\uparrow}-f_k^{\downarrow}\right)^2}{2}}$.  Minimizing $E$ with respect to $p$ or $q$ gives the same equation,
\begin{align}
H \delta \sin \theta \left[ q^2 -p^2 \right]+2 p q \left[H \cos \theta +\sum_k \lambda_k (f_k^{\uparrow}+f_k^{\downarrow})+\sum_k \omega_k ( (f_k^{\uparrow})^2 -(f_k^{\downarrow})^2) \right]=0
\label{minimization_alpha}
\end{align}
\end{widetext}
Solving self-consistently the set of equations determined by Eqs. (\ref{minimization_f_k_g_k_0}--\ref{minimization_alpha}) allows to compute $p$ and $q$ and their evolution with respect to $\theta$ for different values of $\alpha$. The behaviour of the spectral function $J$ notably enters into account through the renormalization factor $\delta$. We recover the non-dissipative values, $p=\cos \theta/2$ and $q=\sin \theta/2$ at $\alpha=0$. From the value of $p$ and $q$, one may compute the Berry curvature and we show its evolution with respect to $\theta$ for different values of $\alpha \leq 1/2$ in Fig. \ref{Berry_curvature_shifted_oscillators} of the main text.

\section{Proof of Eq. (\ref{curvature_sigma_y}) using time-dependent perturbation theory}
\label{appendix_perturbation_theory}
 Let us call $|e_t\rangle$ and $|g_t\rangle$ the excited and ground state of the system at time $t$, associated with the eigenenergies $E_e (t)$ and $E_g (t)$. We project the wavefunction of the system at time $t$ on this instantaneous basis,
\begin{align}
|\psi(t)\rangle=a_g (t) |g_t\rangle+a_e (t) |e_t\rangle.
\label{wavefunction}
\end{align} 
\begin{widetext}
We first show that the non adiabatic response of the system will lead to a non-zero expectation value for $\langle \sigma^y(t) \rangle$. Then we compute $a_e (t)$ and $a_g (t)$.
Let us define $A(\phi)=\langle \partial_{\phi} \mathcal{H}_{TLS} \rangle$. We have 
\begin{align}
A(\phi)&=\langle \psi(t)| \partial_{\phi} \mathcal{H}_{TLS}| \psi(t) \rangle \notag \\
&=|a_g(t)|^2 \underbrace{\langle g_t | \partial_{\phi} \mathcal{H}_{TLS}| g_t \rangle}_{=0}+ |a_e(t)|^2 \underbrace{\langle e_t | \partial_{\phi} \mathcal{H}_{TLS}| e_t \rangle}_{=0}+ a_g(t) a_e^*(t) \langle e_t |\partial_{\phi} \mathcal{H}_{TLS} | g_t \rangle +  a_g^*(t) a_e(t) \langle g_t |\partial_{\phi} \mathcal{H}_{TLS} | e_t \rangle.
\end{align}
Then 
\begin{align}
A(\phi=0)&= H/2 \sin \theta (t) \langle \sigma^y \rangle= \left[ a_g(t) a_e^*(t) \langle e_t |\partial_{\phi} \mathcal{H}_{TLS} | g_t \rangle +  a_g^*(t) a_e(t) \langle g_t |\partial_{\phi} \mathcal{H}_{TLS} | e_t \rangle\right] (\phi=0).
\label{A_0}
\end{align}
It is clear from Eq. (\ref{A_0}) that $\langle \sigma^y (t) \rangle$ is linked to the non-adiabatic response of the system.
To find the time evolution of $a_e(t)$ and $a_g(t)$, we use time dependent perturbation theory, following Refs.~\onlinecite{polkovnikov:course,polkovnikov:PNAS}. Inserting expression (\ref{wavefunction}) into the Schr\"dinger equation and projecting on the state $|e_t \rangle$, we get: \label{perturbation_theory}
\begin{align}
i \dot{a_e}(t) + a_g(t) \langle e_t | \partial_t | g_t \rangle=a_e(t) E_e(t).
\end{align}
Next, we define $\alpha_i (t)=a_i (t) e^{ i \Theta_i (t)}$, with $\Theta_i (t)=\int_{t_0}^t E_i (\tau) d\tau$ for $i=(g,e)$. At first order in $v/G$, we get \cite{polkovnikov:course},
\begin{align}
\alpha_e (t)=i \int_{t_0}^t d\tau \langle e_{\tau} | \partial_{\tau} | g_{\tau} \rangle \exp \left[i \left(\Theta_e (\tau)-\Theta_g (\tau) \right) \right]+o(v/G).
\end{align}
Using integration rules on fast oscillating functions\cite{polkovnikov:course}, one finaly reaches in the case of an initial adiabatic evolution
\begin{align}
a_e (t)=-i v \frac{\langle e_t | \partial_{\theta} \mathcal{H}_{TLS} | g_t \rangle}{\left[E_e(t)-E_g(t)\right]^2}+o(v/G).
\label{excited_coeff}
\end{align}
Inserting the expression (\ref{excited_coeff}) into Eq. (\ref{A_0}), we get:
\begin{align}
\langle \psi(t)| \partial_{\phi} \mathcal{H}_{TLS}| \psi(t) \rangle=-iv\frac{\langle g_t | \partial_{\phi}\mathcal{H}_{TLS} | e_t \rangle\langle e_t | \partial_{\theta}\mathcal{H}_{TLS} | g_t \rangle-\langle g_t | \partial_{\theta}\mathcal{H}_{TLS} | e_t \rangle\langle e_t | \partial_{\phi}\mathcal{H}_{TLS} | g_t \rangle}{\left[E_e(t)-E_g(t)\right]^2}+o(v/G).
\label{sigma_y_curvature}
\end{align}
\end{widetext}
We recognize on the right hand side of Eq. (\ref{sigma_y_curvature}) the expression of the Berry curvature, and we find back Eq. (\ref{curvature_sigma_y}). For $H_0=0$, we have $\mathcal{F}_{\phi \theta}^0=1/2 \sin \theta$. When the evolution is not initially adiabatic, unimportant oscillations appear on top of the Berry curvature signal, as shown in Refs. \onlinecite{polkovnikov:PNAS,polkovnikov:course,Schroer:PRL}. 

The derivation above remains in fact valid for a gapless system \cite{polkovnikov:PNAS}, and a more general criterion for the validity of Eq. (\ref{curvature_sigma_y}) of the main text is $1-|a_g(t)|^2 \ll 1$. This inequality is guaranteed in the presence of a large gap, but it is also fulfilled in gapless systems when there are few excitations in the system. This general criterion can be fulfilled for the ohmic spinboson model by reducing the velocity, until values of $\alpha$ close to $1/2$ (see main text).

\section{SSE Method}
\label{appendix_SSE}
\begin{figure}[tb]
\center
\includegraphics[width=.9\linewidth]{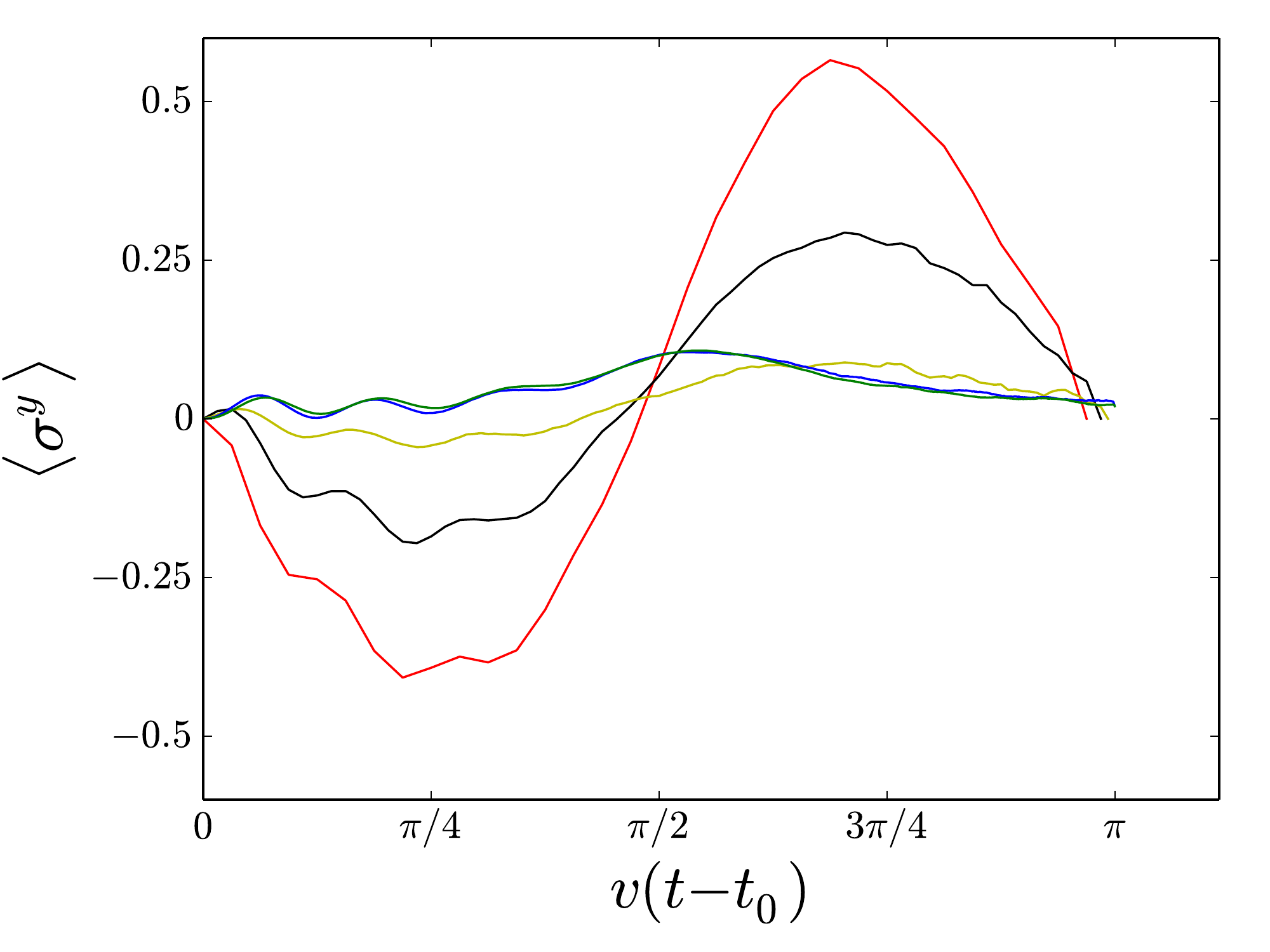}
\caption{Evolution of $\langle \sigma^y \rangle$ with respect to time at $\alpha=0.1$ for $v/H=0.06$, $H/\omega_c=0.01$, and $p=5$ (red), $p=6$ (black), $p=7$ (yellow), $p=8$ (blue) and $p=10$ (green). For greater values of $p$, the curve remains unchanged.}
\label{fig:discretization}
\end{figure}
The SSE method allows to compute the time-evolution of the spin-reduced density matrix $\rho_{\mathcal{S}}(t)$ for $t \geq t_0$, where $t_0$ denotes the initial time. This approach assumes factorizing initial conditions $\rho (t_0)=\rho_B(t_0)\otimes \rho_S(t_0)$, with the bath in a thermal state at inverse temperature $\beta$. In the main text, we take the zero temperature limit $\beta\to \infty$.  Under this factorization assumption, we can show that the elements of the spin-reduced density matrix at time $t \geq t_0$ evolve according to a Stochastic Schr\"{o}dinger-like differential equation (\ref{SSE}), 
\begin{equation} 
i \partial_t | \Phi \rangle = V (t) | \Phi \rangle,
\label{SSE}
\end{equation}
where $ | \Phi \rangle$ is a four-dimensional vector whose components correspond to the elements of $\rho_{\mathcal{S}}$. In Eq. (\ref{SSE}), we have
\begin{equation}
V= \frac{-H \sin \theta(t) }{2} \left( \begin{array}{cccc}
0&e^{-h}&-e^{h}&0 \\
e^{i\pi \alpha}e^{h}&0&0&-e^{-i\pi \alpha}e^{h}\\
-e^{-i\pi \alpha}e^{-h}&0&0&e^{i\pi \alpha}e^{-h}\\
0&-e^{-h}&e^{h}&0
\end{array} \right).
\label{eq:spin_hamiltonian}
\end{equation}
This effective Hamiltonian for $\rho_{\mathcal{S}}$ describes quantum jumps between the different states of the density matrix. The non-zero transition elements are dressed by a time-dependent field $h=h_1+h_2$ which contains a deterministic part $h_1$ depending on the field along the z-direction $h_1=-i \int_{t_0}^t ds (H \cos \theta(s) +H) $ and a stochastic part $h_2$. $h_2$ is more precisely a gaussian random field whose correlations are determined by the bath
\begin{align}
 \overline{ h_2(t) h_2(s)} = & \frac{1}{\pi} Q_2(t-s) + l, \label{height_1} 
\end{align}
where the overline denotes a stochastic average, $l$ is an arbitrary constant and we have at zero temperature
\begin{align}
  Q_2(t)&=\int_0^{\infty} d\omega\frac{J(\omega)}{\omega^2}\left(1-\cos \omega t\right).\label{q2}
\end{align} 
The dynamics is then described by a stochastic process inside the Bloch sphere, whose characteristics depend on the spectral properties of the bath. To recover the spin density matrix, one averages over the stochastic noise. We have more precisely
\begin{equation} 
\left[\rho_S (t)\right]_{ij}=\overline{\langle \Sigma_{ij}| \Phi(t)\rangle},
\label{solution_density_matrix}
\end{equation}
where $| \Phi\rangle$ is the four-dimensional vector solution of Eq. (\ref{SSE}) with initial condition $| \Phi (t_0) \rangle= \left(\left[\rho_S(t_0)\right]_{11} ,\left[\rho_S(t_0)\right]_{12} e^{h(t_0)}, \left[\rho_S(t_0)\right]_{21} e^{-h(t_0)},\left[\rho_S(t_0)\right]_{22} \right)^T$. Vectors $\langle \Sigma_{ij}|$ read $\langle \Sigma_{11}|=(1,0,0,0)$; $\langle \Sigma_{12}|=(0,e^{-h(t)},0,0)$; $\langle \Sigma_{21}|=(0,0,e^{h(t)},0)$; $\langle \Sigma_{22}|=(0,0,0,1)$.

\begin{figure*}[tb]
\center
\includegraphics[width=.36\textwidth]{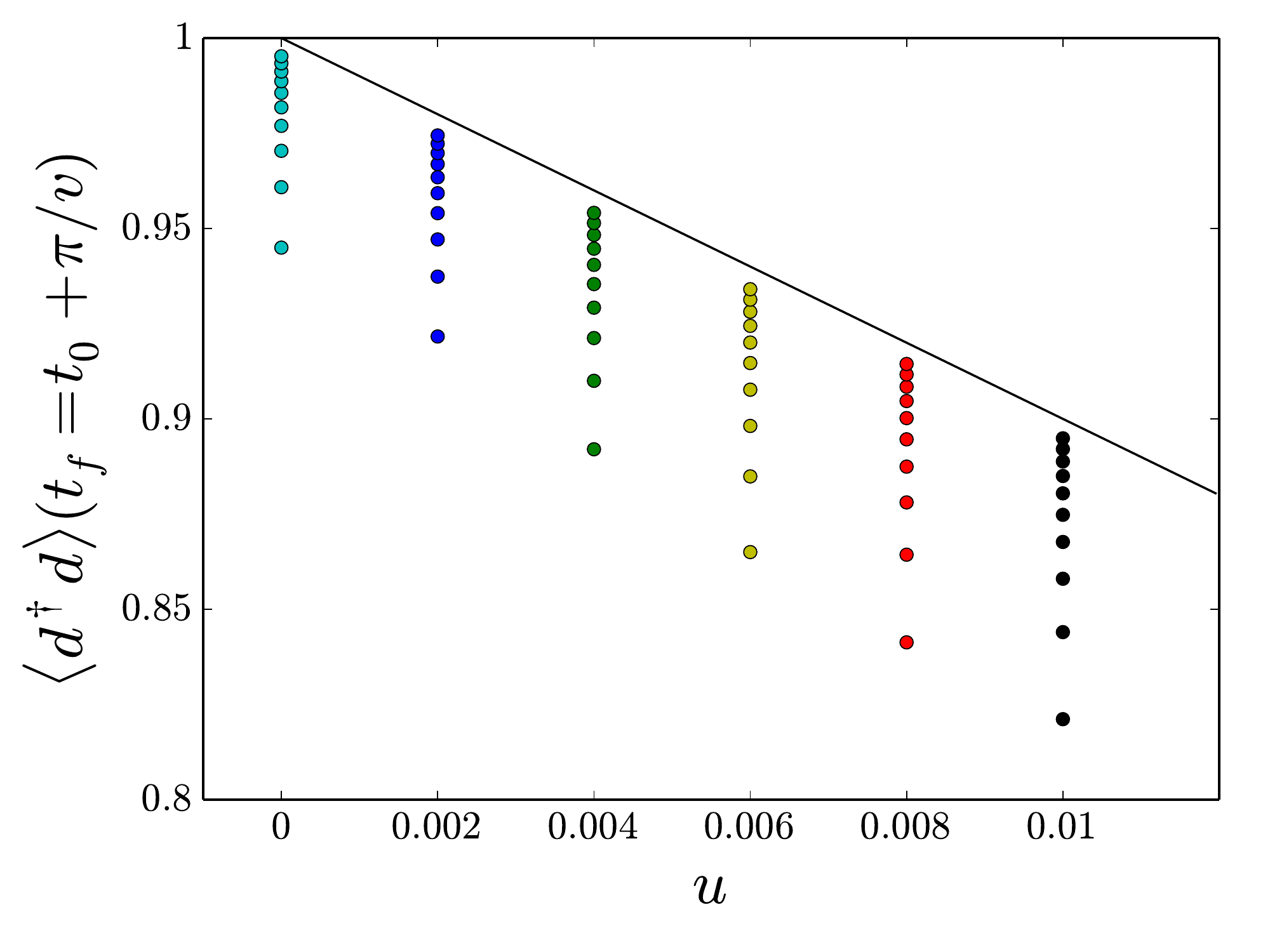}  
\includegraphics[width=.36\textwidth]{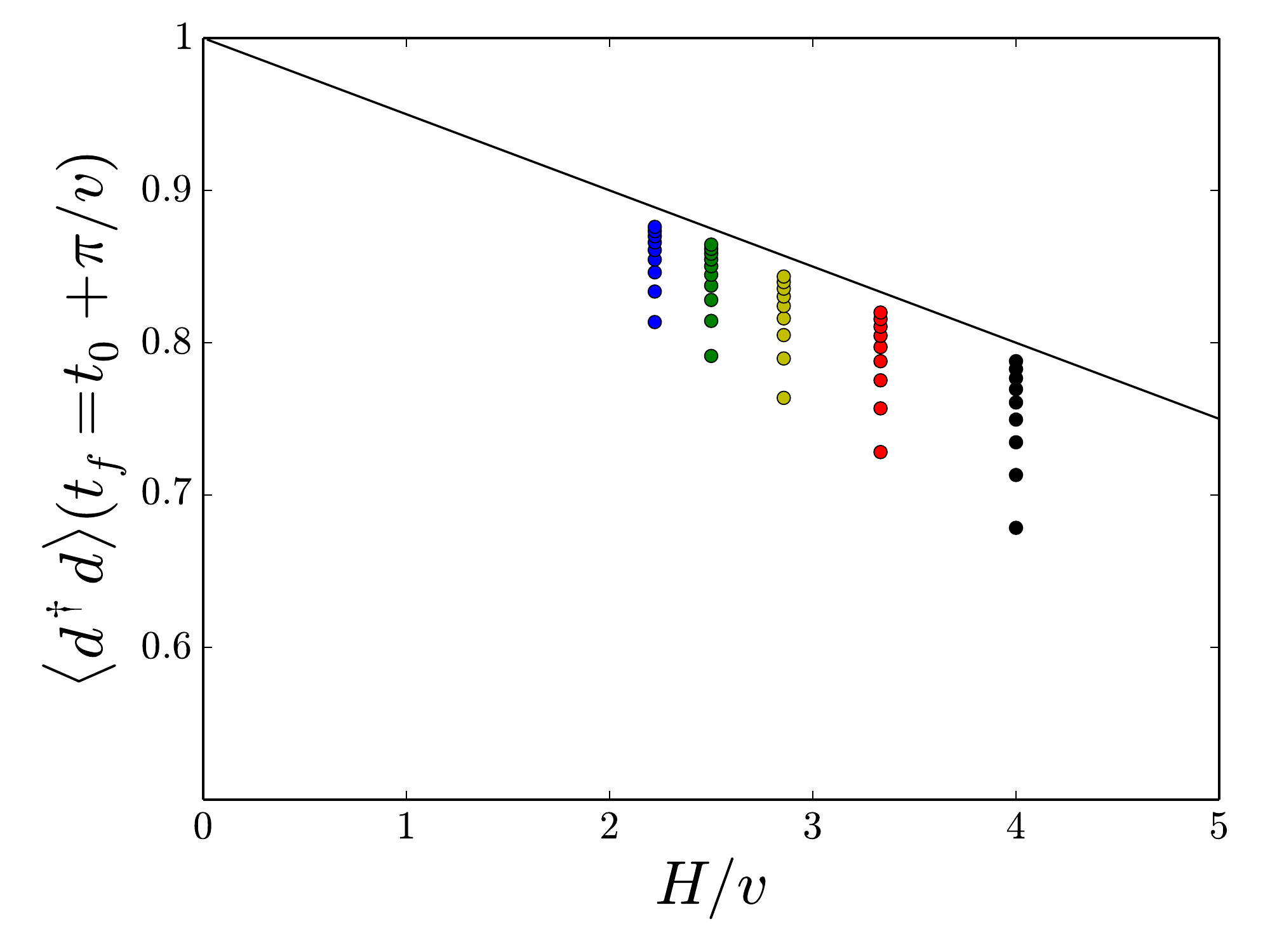}  
\caption{Results from the dynamics using Keldysh formalism in the scaling regime. Left panel: Evolution of $\langle d^{\dagger} d \rangle(t_f=t_0+\pi/v)$ with respect to $u$ for a fixed value of $v/H=0.3$. Right panel: Evolution of $\langle d^{\dagger} d \rangle(t_f=t_0+\pi/v)$ with respect to $H/v$ for a fixed value of $u=0.008$. Each point corresponds to a given value of $\omega_c/H$, regularly spaced from $\omega_c/H=500$ to $\omega_c/H=5000$. The black lines correspond to linear fits. }
\label{mean_field_appendix}
\end{figure*} 

The SSE method is a numerically exact method, whose derivation is based on different results related to Refs. \onlinecite{FV,leggett:RMP,weiss,Stockburger_Mac,Stockburger,2010stoch,stochastic,Rabi_article,Ohmic_systems_article} and can be decomposed into three consecutive steps:
\begin{itemize}
\item Integration of the bosonic degrees of freedom in a path integral formalism \cite{FV}. This integration induces spin-spin interactions, which are long range in time.
\item Rewriting of the spin path in the language of ``Blips'' and ``Sojourns'', following the work of Ref. \onlinecite{leggett:RMP}.
\item  Stochastic unravelling of the bath-induced spin-spin interaction thanks to the introduction of stochastic degrees of freedom \cite{2010stoch,stochastic,Rabi_article,Ohmic_systems_article}. 
\end{itemize}
In practice, the use of the SSE method requires a large number of noise samplings. For each sampling, we solve Eq. (\ref{SSE}) and the spin density matrix is obtained after averaging over the results. Due to the translational invariance of correlations (\ref{height_1}), we use Fourier series decomposition to sample field $h_2$. The use of Fast Fourier Transform algorithm relates Fourier and real time discretization, so that we only have one control parameter $N=2^p$ corresponding to the discretization. For eash simulation, we progressively increase $p$ until the output no longer depends on the discretization, see Fig. \ref{fig:discretization}.

 The SSE method notably gives reliable results for the ohmic spin-boson model in the scaling limit, $H/\omega_c \ll 1$ and $0\leq \alpha \leq 1/2$, as shown in Refs. \onlinecite{2010stoch,stochastic,Rabi_article,Ohmic_systems_article}.
 
 \section{Keldysh approach and deviation from Toulouse point}
 \label{appendix_keldysh} 
 Deviations from the exact mapping point $\alpha=1/2$ result in a additional term $\mathcal{H}_t$ in $\mathcal{H}_T$ of the form\cite{Guinea_bosonization}
\begin{align}
\mathcal{H}_t=U\sum_{k,k'} \left( c^{\dagger}_k c_{k'}-\frac{1}{2}\right) \left( d^{\dagger} d-\frac{1}{2}\right),
\label{additional_term_appendix}
\end{align}
where $U$ is proportional to $u=(1/2-\alpha)$ at first order $u$. This lead-dot interaction term prevents the closure of the equations of motion for the Green functions. To go further, we suppose that the effect of such term can be captured by a mean-field approximation. We develop the bracket in Eq. (\ref{additional_term_appendix}), which gives two different terms. The first one $\mathcal{H}^1_t$ corresponds to an additional field applied on the dot, of strength proportional to $-U$. The second one $\mathcal{H}^2_t$ can be written in the following form,
 \begin{align}
\mathcal{H}^2_t=-\frac{U}{2}\sum_{k,k'} \left( d c^{\dagger}_k c_{k'} d^{\dagger}+ c_{k'} d^{\dagger} d c^{\dagger}_k \right).
\label{additional_term_2_appendix}
\end{align}
We apply then a mean-field approximation on this term, $<AB>=A<B>+<A>B-<A><B>$, with $A,B \in ( d c^{\dagger}_k,c_{k'} d^{\dagger})$. Computing the equations of motion for the retarded dot Green function and the retarded mixed Green function lead to a renormalization of $V(t)$ respectively in $V(t)+U\sum_k <d c^{\dagger}_{k}>(t)$ and $V(t)+U\sum_{k'} <c_{k'} d^{\dagger}>(t)$. This allows then to solve the dynamics at first order in $u$, and we show an example of the numerical results in Fig. \ref{mean_field_appendix}. We find Eq. (\ref{eq:8}) of the main text, with $a=4.1\pm 0.1$ here (we did not keep track of all the constant terms for the numerics).

\bibliographystyle{apsrev4-1}
\bibliography{dissipative_topology_1}

\end{document}